\documentclass[a4paper,12pt]{article}
\usepackage{bm}
\usepackage{amsmath}
\usepackage{amssymb}
\usepackage[dvips]{graphicx}
\usepackage{pdflscape}
\usepackage{multirow}
\usepackage[sort]{natbib}
\usepackage{soul}
\usepackage{xr}
\usepackage{amsmath}
\usepackage{graphicx}
\usepackage{enumerate}
\usepackage{natbib}
\usepackage[hyphens]{url} 
\usepackage[utf8]{inputenc}
\usepackage{mathtools}
\usepackage{amsbsy}
\usepackage{amssymb}
\usepackage{color}
\usepackage{ulem}
\usepackage{placeins}
\usepackage{booktabs}
\usepackage{multirow}
\usepackage{hyperref}
 \usepackage{booktabs,subcaption,amsfonts,dcolumn}
\usepackage{xcolor,colortbl}
\usepackage{subcaption}
\usepackage{verbatim}
\usepackage{chngcntr}
\usepackage{enumitem}

\usepackage{amsthm}

\setlist[enumerate]{leftmargin=*}
\definecolor{Gray}{gray}{0.85}
\definecolor{LightCyan}{rgb}{0.88,1,1}
 
\newcolumntype{b}{>{\columncolor{Gray}}c}
\newcolumntype{a}{>{\columncolor{red}}c}
\newcolumntype{g}{>{\columncolor{green}}c}

\title{ Fisher's   Pioneering work  on Discriminant Analysis and its Impact on  AI \footnote{\bf Based on  ``The  40th  Fisher Memorial Lecture'' delivered  in November 2022, Oxford. Fisher  died 60 years ago and this paper marks the anniversary of his death. For a link to the talk, $http://www.senns.uk/FisherWeb.html$ }
}

\author{ Kanti V. Mardia\\
University of Leeds and University of Oxford.}
\begin{document}
 \date{} 
 \maketitle
\begin{abstract}

 Fisher opened many new areas in Multivariate Analysis, and the one   which we will consider is discriminant analysis. 
 Several papers by Fisher and others followed from his seminal paper in 1936 where he coined the name discrimination function. Historically, his four papers on discriminant analysis during 1936-1940 connect to the contemporaneous pioneering work of Hotelling and Mahalanobis.  We revisit the famous iris data which Fisher used in his 1936 paper and in particular, test the hypothesis of multivariate normality for the data which he assumed.   Fisher constructed his genetic discriminant motivated by this application and we provide a deeper insight into this construction; however, this construction has not been well understood as far as we know. We also indicate how the subject has developed along with the computer revolution, noting newer methods to carry out discriminant analysis, such as kernel classifiers, classification trees, support vector machines, neural networks, and deep learning. Overall, with computational power, the whole subject of Multivariate Analysis has changed its emphasis but the impact of this  Fisher's pioneering work continues  as an integral part  of  supervised learning  in Artificial Intelligence.

\end{abstract}
 

\section{ Introduction}\label{Topic2Intro}

We begin with a quotation  from \citet{Rao1964};  
  Rao was one of Fisher's  PhD students.  He  has
summarised succinctly  the pioneering work of  Fisher in Multivariate Analysis as follows: 

{\it ``Not much was known in the field of Multivariate Analysis by way of theory or applications before Fisher did his pioneering researches. Many important contributions by others have also been inspired by his work. 
 He has forged a number of important tools and demonstrated their use in applied research, specially in the field of biology, all of which are 'bound to stay on the books and be used continuously'. Truly he was the architect of Multivariate Analysis ..." }

 Modern computational advances have gone on to provide many extensions of Fisher's work in Multivariate Analysis. We will discuss his pioneering work in  discriminant analysis  and its impact  on statistical learning /AI. 
\citet{fisher1936} paper is  the landmark paper in  discriminant analysis. 
In his collected papers \citet{Fisher1950} includes his commentary on this paper,
{\it ``This was written to embody the working of a practical numerical
example arising in plant taxonomy, in which the concept of a  discriminant
function  seems to be of immediate service."} \\
\\
{\bf Fisher's 1936-1940 papers}\\

Fisher published four articles on statistical discriminant analysis between 1936 and 1940,  namely
\citet{fisher1936},
 \citet{fisher1938},
 \citet{fisher1939}, and
 \citet{fisher1940}.
 In the first of these (\citet{fisher1936}), he introduced  and applied the Linear
Discriminant Function (LDF) (the Fisher's Rule);
Fisher also considered the problem of  classification and its connection with
regression analysis, including introduction of dummy variables.
\citet{fisher1936} dealt mostly with results needed for the comparison of  three species of Iris, illustrated on the now-celebrated iris data. 
The question posed by Fisher was 
 \\
{\it ``What linear function of the four measurements 
$$  X= \lambda_1 x_1 + \lambda_2 x_2  + \lambda_3 x_3  + \lambda_4 x_4 $$
will maximize the ratio of the difference between the specific means to the standard deviation within species?" }
and then  it says in the paper  {\bf ``The particular linear function which best discriminates the two species will be one for...''},\\
and the subject of Discriminant  Analysis/ Statistical Learning was  borne.
\citet{fisher1938} reviewed his 1936 work and relates it to  Hotelling's $T^2$,  Mahalanobis $D^2$ and Hotelling's Canonical Analysis. Also, he  corrected  his earlier significance test, pointing out
 differences with the usual normal-theory regression. 
 The second half of this paper contains an attempt to extend the earlier work
to the case of $s > 2$  populations. 
 The treatment is somewhat incomplete in \citet{fisher1938}, but is completed 
in \citet{fisher1939} and
 \citet{fisher1940} and in his collected papers \citet{Fisher1950}. \\
 \\
 {\bf Iris Data} \\

\citet{fisher1936} published the full iris data, for the first time.  The data were collected by
  E. \citet{Anderson1935}, with further details in E. \citet{Anderson1936},  but the full data published  by Fisher
is not included in either of the two papers by E. Anderson. The iris data has $p=4$ variables \\

$x_1$ = sepal length, $x_2$ = sepal width,

$x_3$ = petal length, $x_4$ = petal width,\\
\\
and sample size $n=50$ from each of $s=3$ populations, three  species of Iris
$$ \Pi_1: setosa ,  \Pi_2: versicolor, \Pi_3 : virginica ; $$
 the measurements are in cm.
 For the data, see for example, \citet{fisher1936}, \citet{kvm2023},  and  in  R:  library(datasets),
data(iris). Recently, \citet{Unwin2021} have provided more sources on where to find this  data.

 Let $X_1, X_2, X_3$ be the respective $n \times p$ data matrices, where $n=50, p=4$. The sample means are given in  Table \ref {means}.  
 \begin{table}[htbp]
\begin{center}
\begin{tabular}{ll|cccc}
&& \multicolumn{2}{c}{Sepal}& \multicolumn{2}{c}{Petal}\\
&& Length & Width & Length & Width\\
 setosa & $\boldsymbol {\bar {x}} _1$ &   5.006   &     3.428  &      1.462    &    0.246 \\
 versicolor & $\boldsymbol {\bar {x}}_2$ &   5.936    &    2.770   &    4.260   &     1.326 \\
 virginica & $\boldsymbol {\bar {x}}_3$ &  6.588    &   2.974    &   5.552   &     2.026 \\
\end{tabular}
\end{center}
\caption{\label{means} Sample mean vectors  of the  iris data.}
\end{table}

Let  $\boldsymbol{x}_1 ,  \boldsymbol{x}_2$  and  $\boldsymbol{x}_3 $ be the $4 \times 1$ random vectors for the three respective populations   and let  $\boldsymbol{x}_i\sim N(\boldsymbol{\mu}_i, {\Sigma}_i),$  with means and covariance  matrices $\boldsymbol{\mu}_i: 4 \times 1, {\Sigma}_i : 4 \times 4,  \   i=1,2,3, $  respectively. We will indicate if and when the normality assumption is used.  \\
The flowers  of the first two iris species  
 (setosa  and   versicolor)  were taken from   the same natural  colony but the sample of the third iris species  (virginica) is from a different colony.
 One of the aims is to discriminate  the three iris species based on the data.
  \citet{Unwin2021}  seem to have recently traced the source of this data set, and their paper also gives a colour picture of  these three flowers. 
 
 We now outline the contents of this paper which are mostly related to topics in  Fisher's 1936 paper and its subsequent development. Section \ref{Topic2Normality}   assesses  his assumption of multivariate normality for the iris data. In Section  \ref{Topic2CV}, we carry out  discriminant analysis  for all the three iris populations  via canonical variates but the technique of canonical variates arrived after  \citet{fisher1936} which  basically  deals with  two-population discrimination.
New methods have appeared since to provide classification regions and in Section  \ref{Topic2Classify} for  the iris data with two  selected variables, we compare Fisher's discriminant rule  and the related maximum likelihood rule to  two non-parametric methods:  kernel classifier and  classification trees.

 When Fisher wrote his paper of 1936, he had the discriminant method for 2 populations but there were 3 populations in the iris data.  This he handled by going into the so-called genetic 
hypothesis   of  the iris data, which led to his specific  analysis. The construction to handle the hypothesis took him to a new rule  for  discrimination,  but   this construction is  not well understood as far as we know. We give some rationale behind his rule, which we call  ``genetic discriminant'',  in Section \ref{Topic2Genetc}.
In his 1936 paper, Fisher gave an effective way to plot classification histograms of his  genetic discriminant and  in   Section \ref{Topic2Visual}  we examine  some modern multivariate  visualisations   of the various  discriminants.
Matrix algebra is one of the most important mathematical  tools in statistics and especially in multivariate analysis. Fisher used some   matrix algebra in \citet{fisher1936}, but overall matrix algebra came slowly into use; there was some resistance in the 1940's. We give a very brief history on the appearance   of matrix algebra in multivariate analysis   in Section \ref{MatrixH}. 
Finally, Section \ref{Learning} gives a short  overview of how Fisher's discriminant analysis has influenced the field of supervised learning which is an integral part of  AI .

\section{Testing  Multi-normality of the  Iris Data}\label{Topic2Normality}

 Initially,  \citet{fisher1936} sets up the problem without the need to assume any distributions, but for  calculating the  classification error, he says, {\it "We may, therefore, at once conclude that   if the measurements are nearly normally distributed ..."}.  Subsequently, for testing genetical hypothesis he again assumes normality.

We now assess this assumption. For this section, we use the notation that    ${\bm x_r, \ r=1, \ldots, n},$   is  a random sample  from a population in $p$ dimensions with the sample mean vector $\bar {\bm x} $ and sample covariance matrix $\bm S$.
 Using the invariant functions
$$
g_{rs} = (\bf  x_r - \bar{\bf x})^T \bf S^{-1} (\bf x_s - \bar{\bf x}),
$$
\citet{Mardia1970} introduced the following multivariate measures of skewness and kurtosis respectively.
$$ b_{1,p} = \frac{1}{n^2} \sum^n_{r,s = 1} g^3_{rs}, \;\;
 b_{2,p} = \frac{1}{n} \sum^n_{r = 1} g^2_{rr}.$$
Some  of the attractive properties of these measures  are:
\begin{enumerate}
\item[(1)] For the univariate case, $b_{1,1} = m^2_3/m^3_2=b^2_1$  and  $b_{2,1}= m_4/m^2_2=b_2$,
 so these reduce to  the standard measures where the $m$'s  are univariate central  moments.
\item[(2)] $b_{1,p}$ depends only on the moments up to the third order,
  whereas $b_{2,p}$  only on the even moments up to the fourth order.
\item[(3)]These measures are invariant under \emph{affine} transformations.
\item [(4)]  For the multivariate normal distribution, the population  measures  $\beta_{1,p}, 
\beta_{2,p} $   have values 
$$\beta_{1,p}=0,\quad \beta_{2,p} = p(p+2).$$  
\item [(5)] Under normality,  $b_{1,p}$ and $b_{2,p}$, have the following asymptotic
distributions
$$U=  \frac{1}{6}nb_{1,p}\sim\chi_f^2\qquad\textrm{where}\qquad f=\frac{1}{6}p(p+1)(p+2), $$
$$ V=\{b_{2,p}-p(p+2)\}/\{8p(p+2)/n\}^{1/2}\sim N(0,1).$$
These statistics  $ U ,V$ are used to test the null hypothesis of multivariate normality for large samples.  
For small samples, \citet{kvm1974}  has given critical values of these statistics.
\item [(6)]  These measures are available  in all the standard statistical packages and are now one of the most popular measures.

\end{enumerate}

\begin{table}[!htb] 
\begin{center}
\begin{tabular}{lcccl}
& setosa & versicolor & virginica \\
skewness & 3.1 (25.7) & 3.0 (25.2) & 3.2 (26.3) \\
kurtosis & 26.5(1.3) & 22.9 (0.6) & 24.3(0.2)\\
\end{tabular}
\end{center}
\caption{ \label{SkKr}  For the iris data  $(p=4, n=50) $,  skewness $b_{1,p}$ and kurtosis $b_{2,p}$
for each of the  species  together with the values of  the test statistics $U$ and  $|V|$ which are  in the  brackets.}
\end{table}
We have under normality,  the population values are   $\beta_{1,4}=0$ and 
$\beta_{2,4} = 24$  so it can be seen from the Table \ref{SkKr} that the species have similar skewness and kurtosis values  as for the multivariate normal.
 Further,  the  $5\%$ value for  $U$ is 31.41 with $f=20$,  and the 
  $5\%$ value   for  $| V |$  is 1.96.  So, it can be seen from  Table \ref{SkKr}  that  skewness and kurtosis values  are not significant at the  $5\%$  level of significance. 
 Hence,  Fisher's assumption of  multivariate normality for the iris data is justified by these measures.  
 
 \section { Canonical Variates  }\label{Topic2CV}  
  Suppose ($ X_1,\ldots, X_s)$ are data  matrices where  $  X_j$  $(n_j \times p)$ represents the data  matrix  a sample of size $n_j$ on $p$  variables from 
 population $\Pi_j$,  $j=1,\ldots,s$.  
The task of discriminant analysis is to allocate a {\em new} observation ${\bm x}$  to
one of these $s$ populations. \\
Fisher's approach  was to look for the {\em linear} function
$\bm a^T\bm x$ 
that maximizes the ratio of the between-groups sums of squares and cross products matrix  $B$  to the within-group sums of 
squares and  cross products matrix $W$; that is, $\bm a$ is the vector which maximizes
\begin{equation}
\label{BW}
 \bm a^T B\bm a/\bm a^T W \bm a, \quad \textrm{subject to}\  \bm a^T W^{-1}\bm a =1.
\end{equation}
It can be shown that the vector $\bm a$  is the eigenvector of $ W^{-1} B$ 
corresponding to the largest eigenvalue, normalised as in the above expression (\ref{BW}); see, for example, \citet{kvm2023}. This  tractable practical solution was  provided by \citet{bry1951}.\\
\\
{\bf Classification Rule I.}
Let  $ \bar{\bm x}_j $ be the $j$th sample mean vector, $j=1,\ldots,s$.  Allocate $\bm x$ to $\Pi_l$ if
$$ |\bm a^T\bm x- a^T\bar{\bm x}_l| <|\bm a^T\bm x-\bm a^T\bar{\bm x}_j|\qquad\textrm{for all}\  j \ne l,\ 
j,l =1,\ldots,s. $$
\\
{\bf Canonical Analysis of Iris Data }\\
\\
In general, $ W^{-1} B$ has $k=\min(p,s-1)$ non-zero eigenvalues, and  its eigenvectors 
define the first, second and subsequent ``canonical variates'';  the first $k$ canonical 
variates  summarize the difference between 
the groups in $k$ dimensions. Note that  the first canonical variable provides   the discriminant for 3 populations.
   \citet{kvm1979} pp.344-345  have given the first two canonical variables for the iris data  as    
 \begin{equation}\label{canoni}
 \ell^T_1= (0.83,1.53,-2.20,2.81),\
 \ell ^T_2=(-0.02,-2.16,0.93,-2.84).
\end{equation}
\begin{figure}[!htb]
\begin{center}
\includegraphics[scale=.65,angle=0]{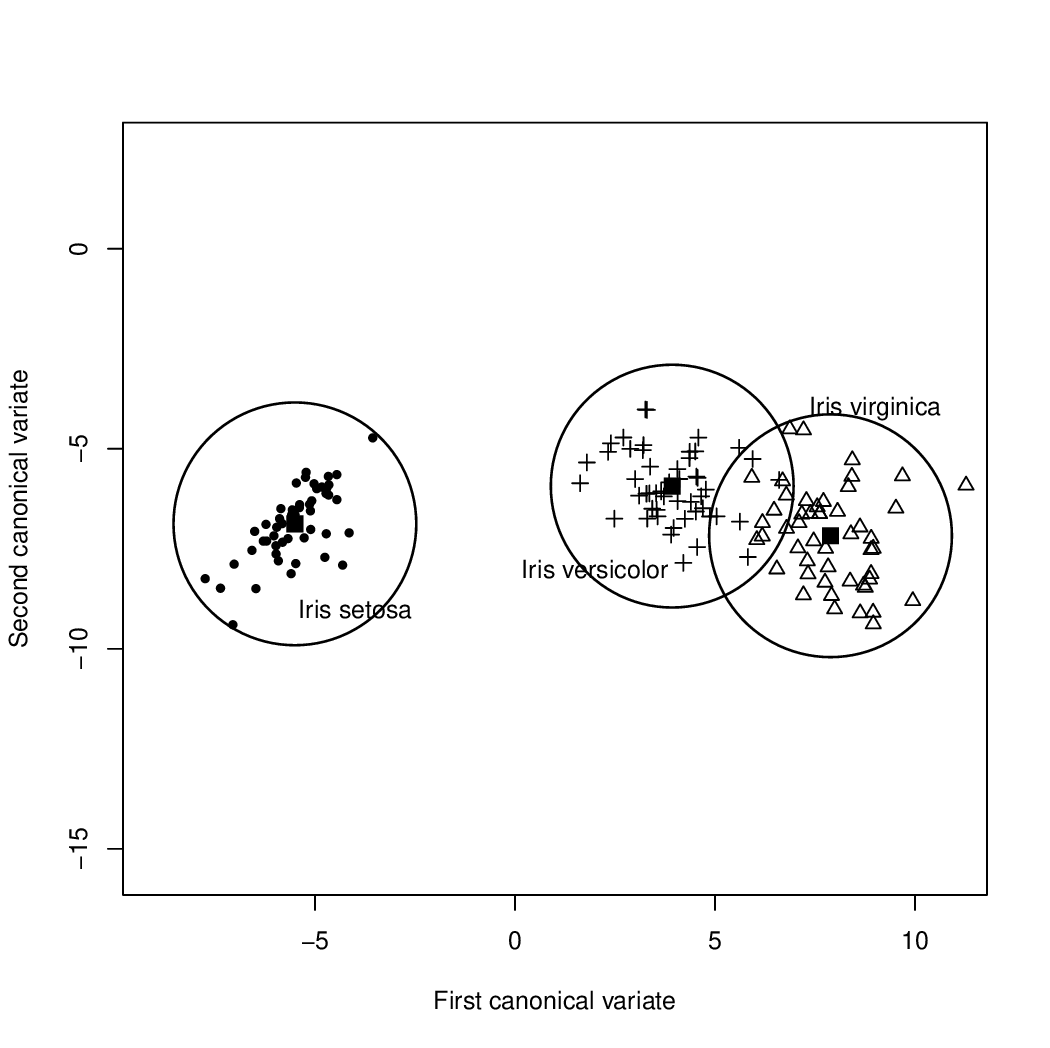}\\
 \end{center}
 \caption{\label{Canonical}The first two canonical variates    of iris data, with approximate 99{\%} provability region   for the
data  with canonical means ( $\blacksquare$). ($\bullet$ setosa, $+$   versicolor and $\triangle$  virginica).}
\end{figure}
Figure (\ref{Canonical}) plots these first two  canonical  variates for the data, with their  99 $\%$  probability region 
  together with their  canonical means. The figure shows that the  three species are different  but there is some  overlap between virginica and versicolor.

 \begin{table}[htbp]
\begin{tabular}{ll|ccc|c}

&& \multicolumn{3}{c}{Actual}\\
&& setosa & versicolor & virginica & Total\\
\multirow{2}{*}{Predicted} & setosa & 50 & 0 & 0 & 50 \\
&  versicolor & 0 & 48 & 0 & 48 \\
&  virginica & 0 & 2 & 50 & 52 \\
& Total & 50 & 50 & 50 & 150 \\
\end{tabular}
\caption{\label{confC}The confusion matrix for the first canonical variable for the full data.}
\end{table} 
 The confusion matrix using only the first canonical variate is given in Table \ref{confC} and the misallocated observations  are  only two, namely, numbers 73 and 84  when  the 150 observations are  ordered row-wise so two versicolor observations are wrongly classified.
           See  Section \ref{Topic2Genetc} for further details.


\section{Classification Regions  for  the iris data with two variables}\label{Topic2Classify}
 Recall, we have  three iris species
 each  with the  sample size $n=50$ and there are four variables $p=4$. For illustrative purpose, we have selected  the first two variables,  $x_1$  sepal length, and    $x_2$  
sepal width. The choice of $p=2$  also helps with 
visualization;  see also Section \ref{Topic2Visual}.
Recall further that the first two iris species  
setosa  and   versicolor  were taken from   the same colony but the sample of the third iris species virginica differ as it was not taken from the same natural colony.

\begin{figure}[!htb]
\begin{center}
\includegraphics[width=0.8\linewidth]{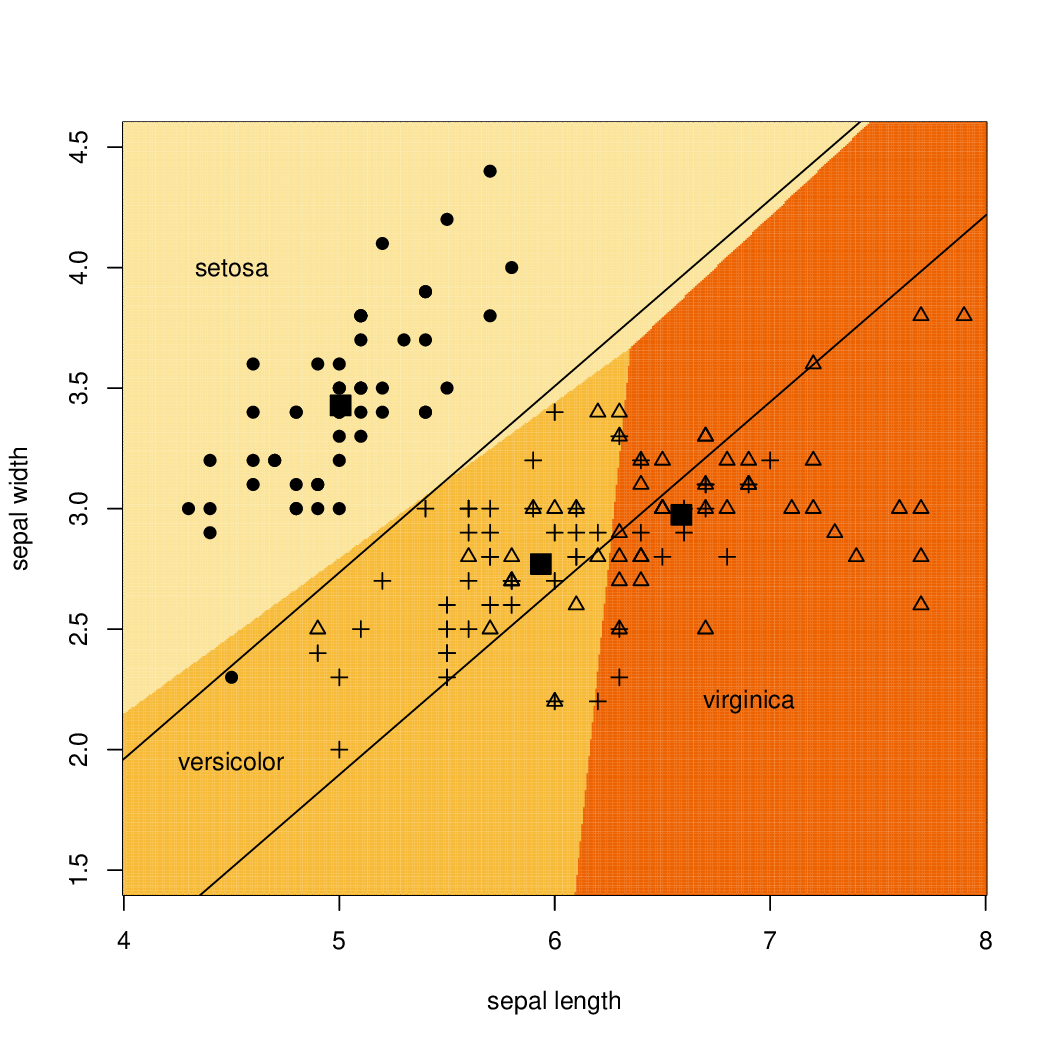}
\end{center}
 \caption{\label{figLinDisc} Classification  by  the Fisher Rule (parallel solid lines as the boundaries) and the ML Rule (boundaries of the three colours). The sample means are denoted by  $\blacksquare$.   The allocation agreement between the two rules is 139 of 150. ($\bullet$ setosa, $+$   versicolor and $\triangle$  virginica).}
\end{figure}

\begin{figure}[!htb]
\begin{center}
\includegraphics[width=0.8\textwidth]{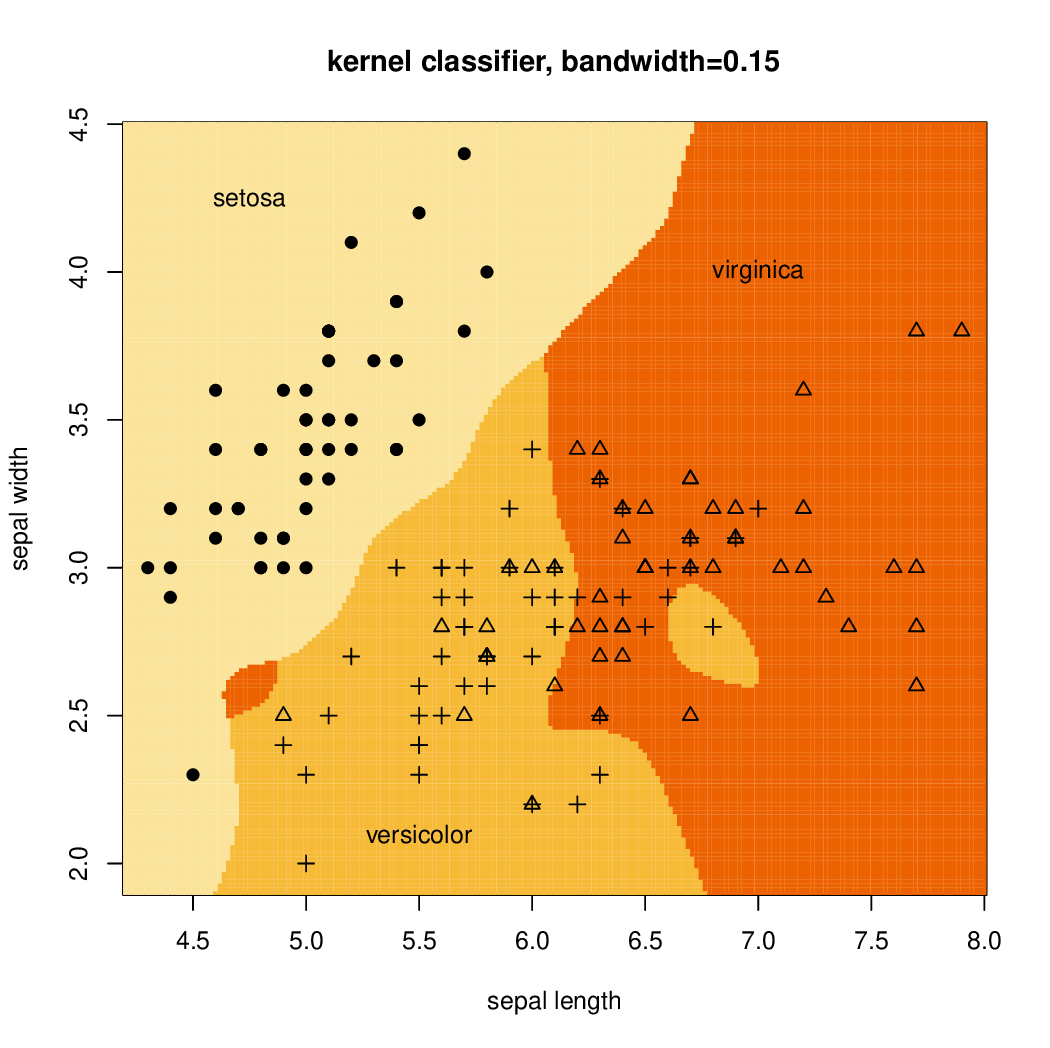}
\end{center}
\caption{\label{figKernel} Kernel classifier. The allocation agreement  of this rule    with the Fisher Rule is 132  of 150  and,   with  the ML Rule is 139  of 150. ($\bullet$ setosa, $+$   versicolor and $\triangle$  virginica).}
\end{figure}

\begin{figure}[!htb]
\begin{center}
\includegraphics[width=0.8\textwidth]{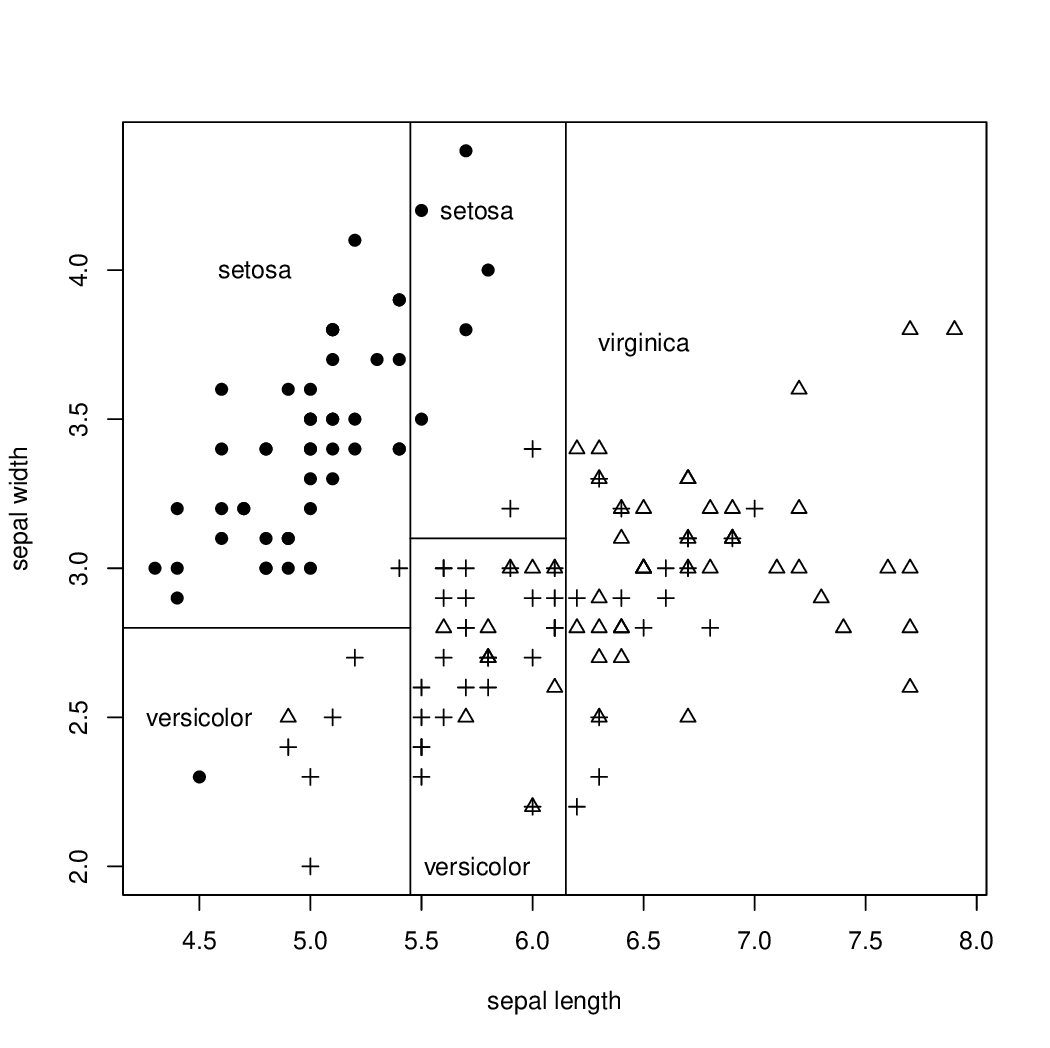}
\end{center}
\caption{\label{figClassTree}  Classification-tree classifier. 
 The allocation agreement  of this rule    with the Fisher Rule is 132  of 150  and,   with  the ML Rule is 141  of 150.
($\bullet$ setosa, $+$   versicolor and $\triangle$  virginica). }
\end{figure}

We give the classification regions, and thus the allocation of each observation to one of three species, for two parametric methods\\ 
   (1) Fisher's linear discriminant  or  the Fisher Rule, Figure \ref{figLinDisc} and \\
   (2)  maximum likelihood discriminant (equal covariances) or the ML Rule,  Figure (\ref{figLinDisc}),\\
  and  two new  statistical learning methods which are non-parametric: \\
    (3) a kernel classifier, Figure \ref{figKernel} and\\ 
       (4) a  classification-tree classifier, Figure \ref{figClassTree}.\\
  
\noindent
\\
\\{\bf Some comments on the iris data}\\
\\
 We have applied  four methods of discrimination/ statistical learning to classify  the iris data (by two variables). The classification regions are similar  though some boundaries are curved rather than straight lines.

 The Fisher Rule  and the kernel classifier  allocations agree for 132 of 150 observations, and the Fisher Rule and classification-tree allocations also agree for 132 of 150.   (The two sets of 132 observations are not identical.).
Whereas, the ML Rule and the kernel classifier  agree  in 139  of 150 observations in allocations,  and the ML Rule and the classification-tree agree in 141  of 150.   Thus the classification methods are   similar on this basis as well.  

The correct classification (prediction) rates are for (1) the  Fisher Rule 117  of 150 (2)  the ML  Rule, 120 of 150, (3) the kernel classifier 127   of 150, and for (4) the classification-tree classifier, 119 of 150.  
These prediction rates are all based on re-substitution.

Thus the method are  very similar on this basis as well.  
In fact,  \citet{Shinmura2016}
  compares 8 different  linear discriminant functions (LDFs) using several different types of data sets including the iris data. The author asserts (p. 53) that  {\it ``Because there are small differences between Fisher's LDF and
other LDFs, we should no longer use iris data as the evaluation data." }
  Thus even using only two variables, the allocation  results are  similar. This is expected since the three samples pass the test of multivariate normality (Section \ref{Topic2Normality}) as under normality the Fisher rule is  optimal when all the variables are used.

\section {Fisher's genetic discriminant }\label{Topic2Genetc}
\subsection {Intoduction} \label{IntGenetic}

\begin{figure}[!htb]
\begin{center}
\includegraphics[scale=.17,angle=0]{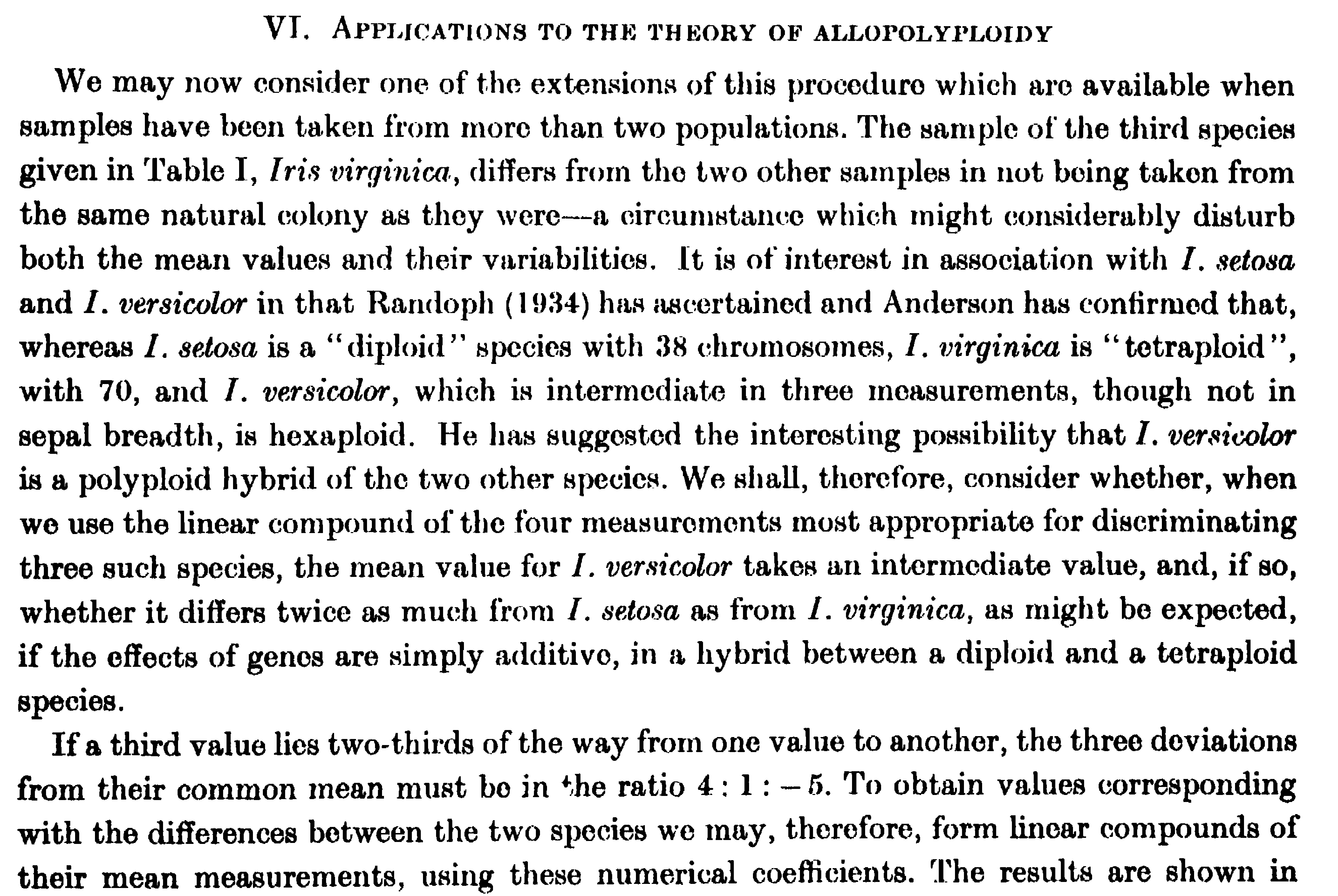}\\
 \end{center}
 \caption{\label{Sec6Text} Reproduced an extract from Section 6 of  \citet{fisher1936} describing reasoning in constructing his genetic discriminant.}
\end{figure}

\citet{fisher1936} in his Section 6 constructed  a  genetic discriminant 
based on  the  reasoning  described in his paper; see Figure \ref{Sec6Text} which reproduces an extract of his  Section 6. He concluded  that 
we  should use the linear  combination of the four measurements most appropriate for discriminating
these 3  species when it is known (from genetic considerations) that  the mean value for versicolor takes an intermediate value  differing twice as much from  setosa as from virginica,
namely, versicolor has been
formed as a 1:2 mixture of setosa and  virginica. 

That is, using our notation  with  the mean vectors    $\boldsymbol{\mu_1}$,  $\boldsymbol{\mu_2}$, and  $\boldsymbol{\mu_3} $ of setosa, versicolor, and virginica respectively,  then
\begin{equation}
\label{Interpoint}
 \boldsymbol{\mu_2}= \frac { \boldsymbol{\mu_1} +2\boldsymbol{\mu_3}}{3}, \quad\text{or} \quad
  \boldsymbol{\mu_1} -3 \boldsymbol{\mu_2} +2\boldsymbol{\mu_3}= 0.
\end{equation}
Fisher's genetic discriminant takes into account this prior information with  the aim   to find the best linear combination of measurements separating
the two putative parents (setosa  and virginica), and then to test whether versicolor lies 1:2 between setosa  and virginica on that combination. (see also,  Figure \ref{AnderGenetic} of E Anderson.)

We now  attempt to justify this  Fisher's genetic  discriminant. Since it pre-specifies a direction of the discriminant, we denote 
 this linear discriminant function (LDF)  in a pre-specified direction by  LDF-PD.
 
\begin{figure}[!htb]
\begin{center}
\includegraphics[scale=.475,angle=0]{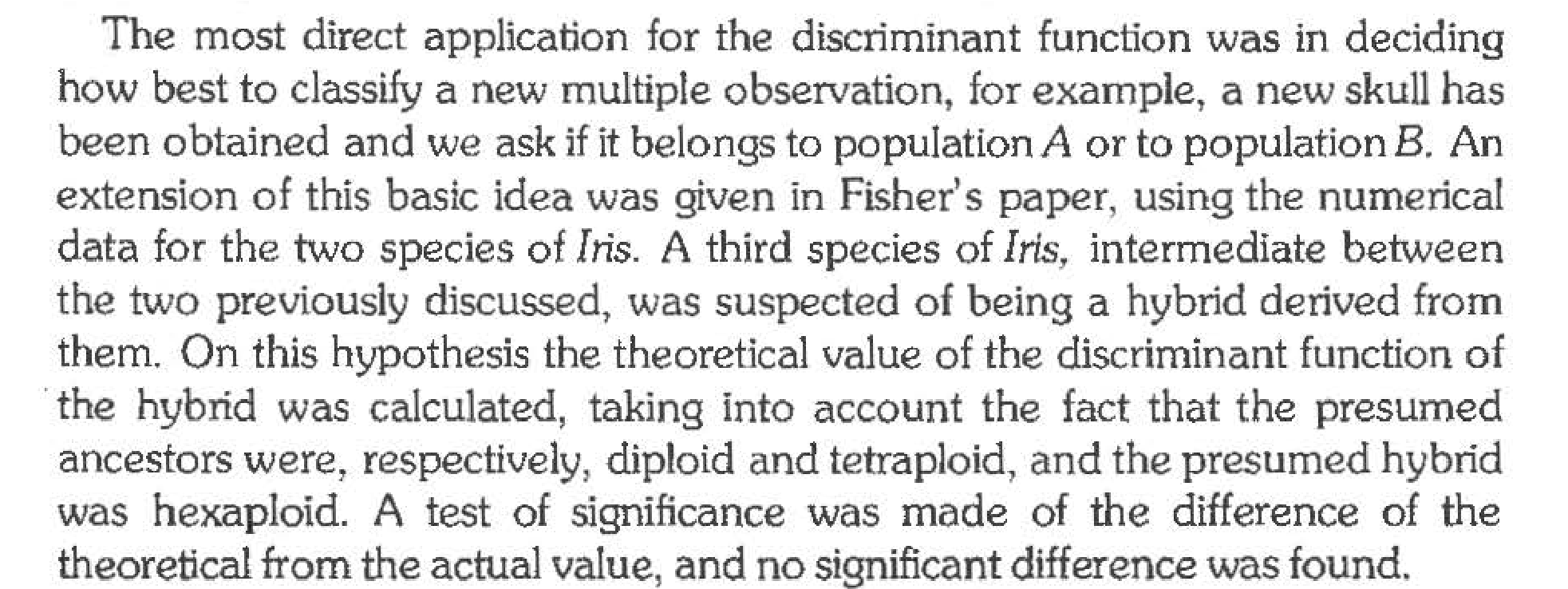}\\
 \end{center}
 \caption{\label{JBox} Excerpt from  \citet{box1978}
  p.333  on a description of how important is  `` Fisher's Genetic Discriminator''.}
\end{figure}

 Recall that when this paper was written, canonical variate analysis was not yet developed to carry out discrimination for more than two populations (3 populations here).
Even then, it is worth finding out how Fisher successfully constructed a plausible  discriminator.
His reasoning is not explicit from the paper and he did not come back  to it in his later writings,
nor do  any later  researchers.
On the other hand, \citet{box1978} points out that how crucial is this work under a pre-specified  direction, see Figure \ref{JBox}.

The relationship of   LDF-PD with  canonical variates is examined in  Section \ref{CanFisher}. 
In Section
\ref{irisData3Pop}, we analyse the iris data using the LDF-PD  and it is interesting to note that the Fisher's genetic discriminant performs as good as the standard canonical variate discriminant; both only 2 observations   misclassifying out of 150 observations. Finally, we formally  test the 
hypothesis  of the genetic  relationship   (\ref{Interpoint}) in Section \ref{Test}.

\subsection {Derivation} \label{IrisDir}

Let us assume that we have $s$ populations  with  the random variables $\bm x_j $  in $p$ dimensions with  population means  and covariances $\boldsymbol{\mu}_j, \Sigma_j, \   j=1,\ldots, s,$  respectively.   We assume that  $\boldsymbol{\mu}_j,   j=1,\ldots, s,$ are all different but {\it collinear}.
In Fisher's derivation  of LDF-PD three  principles are involved and we first give an outline and later on each principle is expanded. 

\begin{enumerate}
\item {\bf Principle 1: Optimal contrast of population  means} Finding   an optimal  contrast  of population means   for  which the groups are most different under  given $s-2$ linear constraints on the (unknown) means.
 That is,  to find the vector    $\boldsymbol{\alpha}=(\alpha_1, \ldots, \alpha_s)^T $  such  that the following linear combination is optimal. 
 \begin{equation}\label{ContrastGen}
 \boldsymbol{\delta}= \sum_{j=1}^s \alpha_j  {\boldsymbol{\mu}_j}, 
\end{equation}
 The optimum criterion is given  below at  (\ref{OptMeans}) with some constraints on $\boldsymbol{\alpha}$. 
 We  denote the optimal   $\boldsymbol{\alpha}$ by $\boldsymbol{\alpha}_0
=(\alpha_1^0, \ldots, \alpha_s^0)^T  $ so that the optimal contrasts on the means  is 
\begin{equation}\label{ContrastZero}
 \boldsymbol{\delta_0}= \sum_{j=1}^s \alpha_j^0  {\boldsymbol{\mu}_j}.
\end{equation}
\item {\bf Principle 2: Population Discriminant}  Using the resulting  $\alpha_j^0$ from Principle 1   form   an ``optimal'' combination of the random variables 
\begin{equation}\label{dDefine}
   \boldsymbol{d}= \sum_{j=1}^s \alpha_j^0  {\boldsymbol{x}_j}.
 \end{equation}
For discrimination,  consider the linear combination  
  \begin{equation}\label{ContrastVec}
 u= \boldsymbol{\lambda}^T\boldsymbol{d}\quad  \text{where} 
 \; \boldsymbol{\lambda}=(\lambda_1, \ldots, \lambda_p)^T, 
\end{equation}
and the optimal   creation  is formed using the signal to noise ratio, namely
\begin {equation} \label{Ratio}
 \frac {(\text{mean)}^2}{\text{variance}}= \frac {(\boldsymbol{\lambda}^T E(\boldsymbol{d}))^2}
  {\boldsymbol{\lambda}^T\{ \text{cov} (\boldsymbol{d}) \} \boldsymbol{\lambda}}. 
 \end{equation} 
Let  the optimal value of  $\boldsymbol{\lambda}$ be $\boldsymbol{\lambda}_0 $. \\
\item {\bf Principle 3: Classification Rule II.}  Let  $\bm x$  be  the new observation  to be allocated.  Using now $\boldsymbol{\lambda}_0 $, calculate  the ``discriminator function''  $u_0 =\boldsymbol{\lambda}_0^T  \bm x$ and allocate $\bm x$ 
 to the  $l$th population $\Pi_ l$ if
\begin{equation}\label{ClassRule}
|\boldsymbol{\lambda}_0^T\bm x -  \boldsymbol{\lambda}_0^T \boldsymbol{\mu}_l | <| \boldsymbol{\lambda}_0^T \bm x- \boldsymbol{\lambda}_0^T \boldsymbol{\mu}_j|    \qquad\textrm{for all}\quad j \ne l,\quad j,l =1,\ldots,s. 
\end{equation}
For this Classification Rule, it  is assumed that all the populations parameters are known.
\end{enumerate}

Note that  these three principles are all related to the populations and not to samples from these populations except in Principle 3 where $\bm x$ is an observation to be allocated. We now work through each of these three Principles in detail.\\

{\bf Principle 1: Optimum Contrasts of Population Means.} 
 We now obtain the  optimum linear combination (\ref{ContrastZero}).
  Let us assume  there are $s-2$ known linear constraints on the means, such as 
 \begin{equation}\label{Constrained}
 \sum_{j=1}^s c_j  {\boldsymbol{\mu}_j} =0
\end{equation}
 where $(c_1, \ldots, c_s)$  are some known constants. Further, we assume (see below for the rationale behind these constraints  for the  particular case of  $s=3$ but the same argument extends for any $s$.)
\begin{equation}\label{alphaScale}
  \sum_{j=1}^s  \alpha_j=0, \sum_{j=1}^s \alpha_j^2=1 .
\end{equation}
We define our  criterion (optimum criterion) to be optimized  is   
 \begin{equation}\label{OptMeans}
 \sum_{j=1}^s ||\alpha_j  (\boldsymbol{\mu}_j-\bar{\boldsymbol{\mu} })||^2
\end{equation}
where $\bar{\boldsymbol{\mu}}$ is the mean of  the $\boldsymbol{\mu}_j$. The criterion  measures the Euclidean distances between the means.
 
 In the discussion, we assume that $s=3$ initially and restrict to  the constrain  (\ref{Interpoint}) used by Fisher  and then give an extension.  (Note there  is only  one constraint in this case and is  given  in  (\ref{Interpoint})). That is,  
  our  linear combination is of the form 
  \begin{equation}
   \bm \delta= \sum_{j=1}^3 \alpha_j \bm \mu_j
 \label{opt3}.
\end{equation}   
and the means are constrained to 
\begin{equation}
\boldsymbol{\mu}_1-3\boldsymbol{\mu}_2+2\boldsymbol{\mu}_3=\boldsymbol{0}\,.
\label{mu}
\end{equation}
 Using the collinearity of the $\boldsymbol{\mu}_j$, we can write
\begin{equation}
\boldsymbol{\mu}_j=\alpha_j^0 \boldsymbol{\beta} + \boldsymbol{\gamma}, \quad
\boldsymbol{\beta}^T  \boldsymbol{\beta}= 1, \quad
\boldsymbol{\alpha}_0 \neq \boldsymbol{0}, \quad
j=1,2,3,
\label{muj}  
\end{equation}
where $\boldsymbol{\beta}$  represents the direction of species differences,
$\boldsymbol{\alpha}_0 = (\alpha_1^0,\alpha_2^0,\alpha_3^0)^T$ is the aforementioned linear combination,
and     $\boldsymbol{\gamma}$ is a common translation vector.  
 Without any loss of generality, we can take
 \begin{equation}
 {\alpha}_1^0+{\alpha}_2^0+{\alpha}_3^0=0, 
 \label{Constraint1}
\end{equation} 
 by absorbing  $\bar{ \alpha}_0 \boldsymbol{\beta}$ into  $ \boldsymbol{\gamma}$  where   $\bar{\alpha}_0= ( { \alpha}_1^0+{ \alpha}_2^0+{ \alpha}_3^0)/3$. 
 Substituting (\ref{muj})  into (\ref{mu}) gives
\begin{equation}
    (\alpha_1^0-3\alpha_2^0+2\alpha_3^0)\boldsymbol{\beta}=0
\end{equation}
or, since $ \boldsymbol{\beta} \neq \boldsymbol{0}$,
\begin{equation}
    \alpha_1^0-3\alpha_2^0+2\alpha_3^0=0.
    \label{Constraint2}
\end{equation}
From the two constraints  (\ref{Constraint1})  and (\ref{Constraint2}), 
$\boldsymbol{\alpha}=(\alpha_1, \alpha_2, \alpha_3)^T $ is given, up to a scaling constant, by 
\begin{equation}\label{alpha0}
    \boldsymbol{\alpha}_0=(-5, 1, 4)^T, \bm \delta _0=
-5\boldsymbol{\mu}_1+\boldsymbol{\mu}_2+4\boldsymbol{\mu}_3.
\end{equation}

We now show that  $\boldsymbol{\alpha}_0$ satisfies the  optimal criterion given at (\ref{OptMeans}). We have 
\begin{equation}
  u (\boldsymbol{\alpha})=   (\sum_{i=1}^3 \alpha_i (\bm \mu_i-\bar{\boldsymbol{\mu}})^T) (\sum_{j=1}^3  \alpha_j (\bm \mu_j-\bar{\boldsymbol{\mu}})), 
 \label{object}
\end{equation}   
where $\sum_{j=1}^3 \alpha_j = 0$.    We have    $\bm \mu_j= \alpha_j^0 \bm \beta$   where  $\bm \beta$ is a  unit vector. Substituting this value of   $\bm \mu_j$ in
  (\ref{object}), we get 
$$ u (\boldsymbol{\alpha})=\sum_{i=1}^3 \sum_{j=1}^3 \alpha_i  \alpha_j \alpha_i ^0\bm\beta^T  \bm \beta \alpha_j^0   =  (\sum_{j=1}^3 \alpha_j  \alpha_j ^0)^2 .$$
Hence  $u (\boldsymbol{\alpha})$ is maximum  when   $\sum_{j=1}^3 \alpha_j  \alpha_j ^0=
\boldsymbol{\alpha}^T \boldsymbol{\alpha}_0$ is maximised so  by  the  Cauchy–Schwarz inequality we have  $\boldsymbol{\alpha} \propto \boldsymbol{\alpha}_0.$

Finally,  we note that  for $s>3$, we need more known constraints, that is $s-2$ constraints  to  get a single optimum direction  where for   $s=2$, the optimal linear combination is trivially 
 $\boldsymbol{\mu}_1  - \boldsymbol{\mu}_2$. \\

 {\bf Principle 2: Optimum Linear Combination of  Random Variables.} 
 Now we assume that the vector  $\boldsymbol{\alpha}_0=(\alpha_1^0, \ldots, \alpha_s^0)^T $  is known and we 
define the    random vector 
\begin{equation}\label{ContrastVec}
 \boldsymbol{d}= \sum_{j=1}^s \alpha_j^0  {\boldsymbol{x}_j},
\end{equation}
where the $\boldsymbol{x}_j$' s are distributed independently with the j{\it th} mean vector  ${\boldsymbol{\mu}_j}$ and the j{\it th} covariance matrix $\Sigma_j$; $j=1,\ldots, s$ .
So that 
\begin{equation}\label{ContrastGen}
 E(\boldsymbol{d})= \sum_{j=1}^s \alpha_j ^0 {\boldsymbol{\mu}_j}, \quad \text{cov} (\boldsymbol{d})  = \sum_{j=1}^s ({\alpha_j ^0})^2 {\Sigma_j}.
\end{equation}

Let $  \boldsymbol{\lambda}^T \boldsymbol{x}$ be the ``discriminant''  function   where $\boldsymbol{x}$ is  a new observation  we define  the  discriminant   
$ \boldsymbol{\lambda}^T \boldsymbol{x}$ where  $\boldsymbol{\lambda}$
 by maximising with respect to 
$\boldsymbol{\lambda}$ the signal to noise ratio given at (\ref{Ratio}) in terms of $\bm d$  so 
   
 \begin{equation}
 \boldsymbol{\lambda} \propto(\sum_{j=1}^s ({\alpha_j ^0})^2\Sigma_j)^{-1} (\sum_{j=1}^s \alpha_j^0  {\boldsymbol{\mu}_j}).
 \label{Frule}
 \end{equation}
A proof of  (\ref{Frule}) on maximising (\ref{Ratio}) can be derived using Corollary A.9.2.2 of \citet{kvm1979}.
Note that  there is no assumption on    $ \text{cov} (\boldsymbol{d})$ but as the $ \boldsymbol{x_j} $`s  
 are independent, so that  $ \text{cov} (\boldsymbol{d})= \sum_{j=1}^s  ({\alpha_j ^0})^2 \ \text{cov} (\boldsymbol{x_j})  $ though this rule does not assume equality of covariances.\\
 
\begin{figure}[!htb]
\begin{center}
\includegraphics[scale=.13,angle=0]{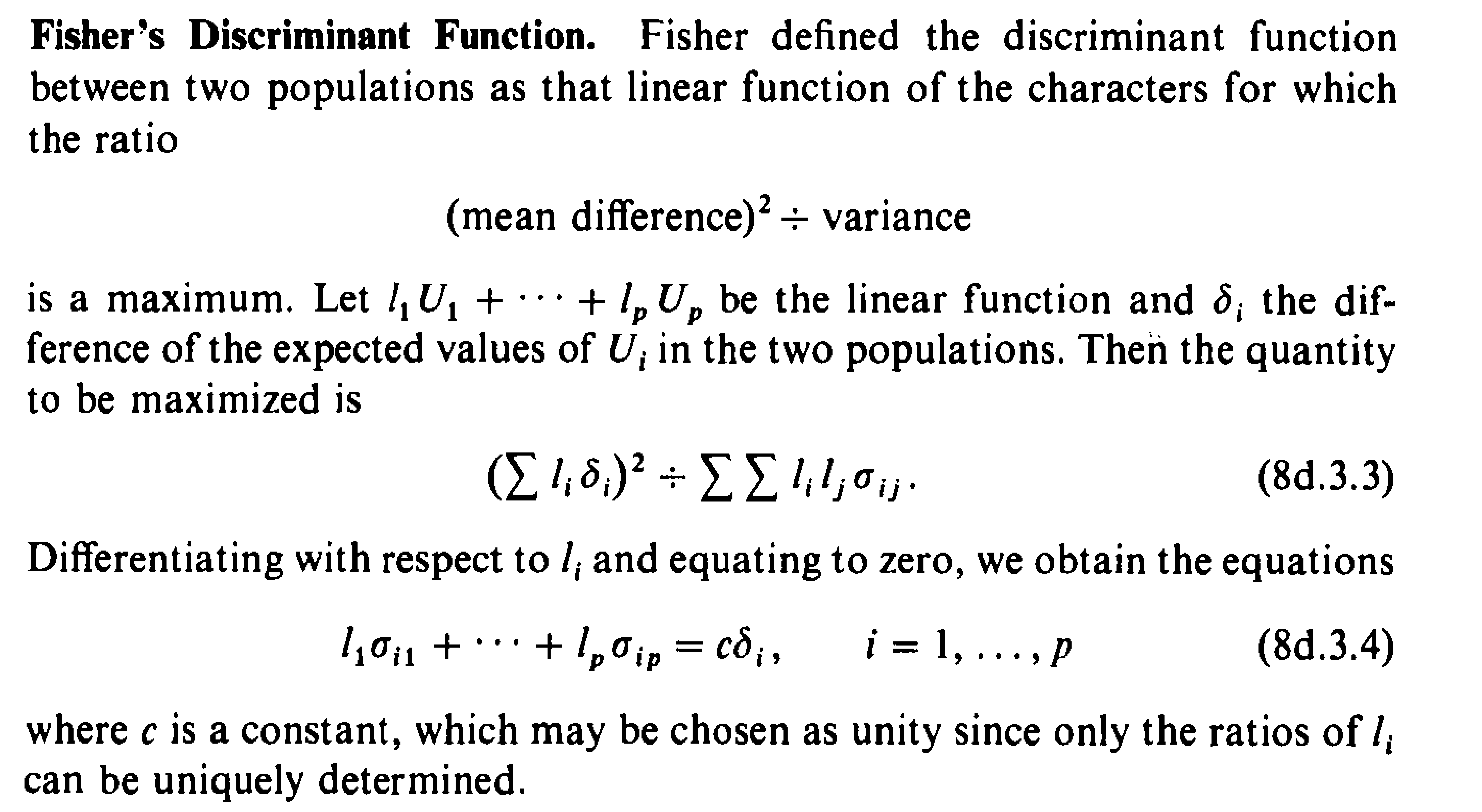}\\
\end{center}
 \caption{ \label{RaoDisc} Extract from \citet{CRrao1973} showing the population discriminant  for two populations.  }
\end{figure}

 The above formulation might look strange as $\boldsymbol{d}$ at  (\ref{dDefine}) is defined as a linear combination of random vectors $\boldsymbol{x}_1,\ldots, \boldsymbol{x}_p$  representing different populations. Further, population discriminant using Fisher's signal to noise ratio is mostly seen in the sample version. However, there are some exceptions. We describe here a particular case which \citet{CRrao1973}  treated and  
Figure \ref{RaoDisc}  gives an extract from  \citet{CRrao1973}  indicating how Fisher's discriminant function can be formulated for two populations. In Rao's case, 
 $s=2, \alpha_1^0=1,  \alpha_2^0=-1$ in  (\ref{dDefine}). He  then used the signal to noise ratio (as we have done in 
(\ref{Ratio}) for our general case) to obtain his $(\l_1, \ldots, \l_p)$ which is equivalent to our 
$ \boldsymbol{\lambda}=(\lambda_1, \ldots, \lambda_p)^T.$  However, it is to be noted that in Rao's case $\alpha$'s are known and  his case is only for $s=2$ what his Book required. 

We restate\\
{\bf  CLASSIFICATION RULE II}
  Allocate $\bm x$ to the  $l$th population $\Pi_ l$ if
\begin{equation}\label{ClassRule}
|\boldsymbol{\lambda}^T\bm x -  \boldsymbol{\lambda}^T \boldsymbol{\mu}_l | <| \boldsymbol{\lambda}^T \bm x- \boldsymbol{\lambda}^T \boldsymbol{\mu}_j|    \qquad\textrm{for all}\quad j \ne l,\quad j,l =1,\ldots,s. 
\end{equation}
Note this rule is standard rule fro the sample case for the canonical variate discrimination and,  for $p=1$, it simply becomes
\begin{equation}\label{ClassRuleUni}
|\bm x -  \boldsymbol{\mu}_l | <|  \bm x-  \boldsymbol{\mu}_j|    \qquad\textrm{for all}\quad j \ne l,\quad j,l =1,\ldots,s. 
\end{equation}
Indeed, so it does not depend on covariance matrices. However, this is different than  the likelihood based rule and these   take into account the differences in covariance matrices. Consider the univariate case with two populations: $\Pi_1$ is   the $N(\mu_1,\sigma_1^2)$ distribution, and 
$\Pi_2$ is  the $N(\mu_2,\sigma_2^2)$  distribution. Let $\mu_2>\mu_1$ and 
$\sigma_1>\sigma_2$. Then maximum  likelihood rule  
allocates new $x$ to $\Pi_1$ if
\begin{equation}\label{UniClassRule}
x^2\left(\frac{1}{\sigma_1^2}-\frac{1}{\sigma_2^2}\right)
-2x\left(\frac{\mu_1}{\sigma_1^2}-\frac{\mu_2}{\sigma_2^2}\right)+
\left(\frac{\mu_1^2}{\sigma_1^2}-\frac{\mu_2^2}{\sigma_2^2}\right)<
2\log\frac{\sigma_2}{\sigma_1}.
\end{equation}
 Thus the rule takes into account that  the $\sigma$'s are different. Further,  the set of the $x$'s for which this inequality
is satisfied forms two distinct regions, one having low values of $x$ and the 
other having high values of $x$. For $\sigma_1=\sigma_2$, the rule is the same as the above Fisher's rule.\\
 {\bf Estimation.} Given a sample, we get the sample discriminant function  following Fisher's idea, by plugging in the standard estimates  $\bar{ \boldsymbol{x}_j}, S_j,$ of 
$\boldsymbol{\mu}_j, \Sigma_j,  j=1,\ldots, s,$, respectively   in  (\ref{Frule}) so we have the coefficient vector 
\begin{equation}
 \boldsymbol{\lambda} \propto(\sum_{j=1}^s  \alpha_j ^2 S_j)^{-1} (\sum_{j=1}^s   \alpha_j   \bar{ {\boldsymbol{x}_j}}).
 \label{FruleS} 
 \end{equation}
 Now we can allocate new observation $\bm x$ using the plugged in version of  (\ref{ClassRule}).
We can extend to the ML rule   under multivariate normality but this will be not pursued here.

\subsection {Relation of  LDF-PD with  Canonical Variates}  \label{CanFisher}
The LDF-PD  given in  Section \ref{IrisDir} raises several questions including how the Classification Rule II relates to canonical variates.  
Here we  give its relationship with canonical variates for a particular case of  $\Sigma_j = \Sigma, \     j=1,\ldots, s.$ 
Let $W_0$ and $B_0$ be the within- and between-population matrices respectively. 
\begin{equation}
    W_0 =  \Sigma,
\end{equation}
\begin{equation}
    B_0 = \boldsymbol{\delta} \boldsymbol{\delta}^T.
 \label{B}    
\end{equation}
For the eigenvector  $\boldsymbol{\lambda}$, we need to solve
 \begin{equation}
 \Sigma^{-1} B_0 \boldsymbol{\lambda}=\beta \boldsymbol{\lambda} 
 \label{eigeneq}
 \end{equation}
where $\beta$ is an eigenvalue. We find that the first canonical variate,
corresponding to the single non-zero eigenvalue $\beta$, is
\begin{equation}
    \boldsymbol{\lambda} \propto \Sigma^{-1} \boldsymbol{\delta}\quad \text{or}\quad
\boldsymbol{\lambda}     \propto \Sigma^{-1} (\sum_{j=1}^s \alpha_j  {\boldsymbol{\mu}_j}).
    \label{Canonic}
\end{equation}
This is the same $\boldsymbol{\lambda}$  as in Fisher's rule given by (\ref{Frule}) when  $\Sigma_j = \Sigma, \     j=1,\ldots, s.$ 
 A proof  of  (\ref{Canonic}) follows on substituting this value of
 $\boldsymbol{\lambda}$ into (\ref{eigeneq}) together with  (\ref{B}).

\subsection {iris data} \label{irisData3Pop}
Now consider our iris  case. As shown above at (\ref{opt3}), the optimal linear combination     $\boldsymbol{\delta}$  and the corresponding random vector  $\boldsymbol{d}$  are given by
\begin{equation}\label{Contrast}
 \boldsymbol{\delta}= -5\boldsymbol{\mu}_1+\boldsymbol{\mu}_2+4\boldsymbol{\mu}_3,\quad
 \boldsymbol{d}= -5\boldsymbol{x}_1+\boldsymbol{x}_2+4\boldsymbol{x}_3.
\end{equation}
That is  substituting  $\alpha_1=-5, \alpha_2 =1, \alpha_3= 4)$  in   (\ref{Frule}),  the ``genetic discriminant''  is given by 
\begin{equation} \label{comb}
   \hat {\boldsymbol{\lambda}}^T \boldsymbol{x}= (-5\boldsymbol{\mu}_1+\boldsymbol{\mu}_2+4\boldsymbol{\mu}_3)^T({25 {\Sigma}_1+ {\Sigma}_2+16 {\Sigma}_3})^{-1}   \boldsymbol{x}.
\end{equation}
The estimates $\bar{ \boldsymbol{x}_j}, S_j$
of $\boldsymbol{\mu}_i $ and $\Sigma_i$, $i=1,2,3$  in  (\ref{comb}), as done in 
(\ref{FruleS}), yield the estimated weight vector $\hat {\boldsymbol{\lambda}}$.
It is found  that $100 \times   \hat {\boldsymbol{\lambda}}^T \boldsymbol{x}$ is the genetic discriminant
\begin{equation}\label{u}
 \hat {u}  = -3.30 x_1 -2.76x_2 +8.87x_3  +9.93x_4.
\end{equation}
Indeed,  \citet{fisher1936} gave the discriminant (\ref{comb}) without any explanation. The confusion matrix  for this discriminant is  shown in Table \ref{confF}. 
Note that the only two errors, out of 150 predictions, are
to misallocate observations 71 and 84  (the numbering follows the order in the Fisher tabulated data) which belong to versicolor and are allocated to virginica.\\

\begin{table}[htbp]
\begin{tabular}{ll|ccc|c}
&& \multicolumn{3}{c}{Actual}\\
&& setosa & versicolor & virginica & Total\\
\multirow{2}{*}{Predicted} & setosa & 50 & 0 & 0 & 50 \\
&  versicolor & 0 & 48 & 0 & 48 \\
&  virginica & 0 & 2 & 50 & 52 \\
& Total & 50 & 50 & 50 & 150\\
\end{tabular}
\caption{\label{confF}The confusion matrix for the genetic discriminant.}
\end{table}
Fisher does not give a confusion matrix and his work was prior to the canonical variates as we have mentioned above in  Section \ref{Topic2CV}.  Note that  these both discriminant have the same confusion matrix but mistakes for observations  73 and 84 by canonical discriminant rather than  71 and 84 by the genetic discriminator and  both from versicolor allocated ``wrongly'' to virginica.\\

In fact, the 
 first canonical variable given at (\ref{canoni}) normalised so the length is unity then it becomes
$$\ell_1^{*T}=(0.209,  0.385, -0.554,  0.708)$$
where as the normalised genetic discriminant is 
$$\lambda^{*T}= (-0.236, -0.197,  0.634, 0.710)$$
 $\cos(\ell_1^*, \lambda^* )= 0.9997$
which shows that  these two are  are very  close and differ due to the original different normalisation.

It can be seen from the visual plots of the data that the covariances for the three populations are different (see also Figure \ref{FFig1})   (can be tested formally by the Box M-test)  but it is interesting that both lead to the same level of error  though  the  canonical variates  rule assume the equality whereas the Fisher genetic discriminate does not.

\begin{table}[htbp]
\begin{tabular}{ll|ccc|c}
&& \multicolumn{3}{c}{Actual}\\
&& setosa & versicolor & virginica & Total\\
\multirow{2}{*}{Predicted} & setosa & 50 & 0 & 0 & 50 \\
&  versicolor & 0 & 48 & 1 & 49 \\
&  virginica & 0 & 2 & 49 & 51 \\
& Total & 50 & 50 & 50 & 150\\
\end{tabular}
\caption{\label{conM} The confusion matrix for the ML discriminant.}
\end{table}

For completeness, we give in Table \ref{conM}, the allocations using the ML discriminant with all the four variables; recall that only two variables were used in Section \ref{Topic2Classify}).
We note from the Table that  3 observations out of 150  are misclassified--   2 versicolor observations  gets allocated  to virginica and, one virginica  to versicolor so the behaviour is very similar. 

{\bf ``Shape''  Discriminants}

Define  $$ x=\text{sepal length/petal length}=x_1/x_3, y=\text{sepal width/petal width}= x_2/x_4. $$
 \citet{Anderson1936} used the non-linear discriminant  defined by    
\begin{equation}\label{Index}
z_A=  x+y  = \frac{x_1}{x_3} + \frac{x_2}{x_4} , 
\end{equation}  
 which  he called  it  an Index for a basis for comparison. This is in some sense a ``shape'' discriminant. 
  \citet{fisher1936}  made no comment on this Index, perhaps as  the two papers came out in the same year.  If we use the E. Anderson Index $z_A$ as a discriminant, the confusion matrix is given in Table \ref{confA1} 
so 11 observations are misclassified, namely   4 observations of  setosa go to versicolor and 3 of virginica go to versicolor.
\begin{table}[htbp]
\begin{tabular}{ll|ccc|c}
&& \multicolumn{3}{c}{Actual}\\
&& setosa & versicolor & virginica & Total\\
\multirow{2}{*}{Predicted} & setosa & 46 & 0 & 0 & 46 \\
&  versicolor & 4 & 46 & 3 & 53 \\
&  virginica & 0 & 4 & 47 & 51 \\
& Total & 50 & 50 & 50 & 150\\
\end{tabular}
\caption{\label{confA1}The confusion matrix for the E. Anderson's Index.}
\end{table}
 \\
 
If we use $(x,y)$ as the variables,  it is found that  the  Fisher  discriminant  is 
 $$ z= 3.58 x +0.06 y $$
 so the length ratio dominates rather than the width ratio.
  The confusion matrix  for the two  shape ratios  is given  in Table \ref{confA2} so there are 8  observations misclassified,   with   7 observations of   versicolor go to virginica and 1  of  virginica goes to  versicolor so but  so it is  a bit better than the E. Anderson's Index (11 observations are misclassified).
\begin{table}[htbp]
\begin{tabular}{ll|ccc|c}
&& \multicolumn{3}{c}{Actual}\\
&& setosa & versicolor & virginica & Total\\
\multirow{2}{*}{Predicted} & setosa & 50 & 0 & 0 & 50 \\
&  versicolor & 0 & 43 & 1 & 44 \\
&  virginica & 0 & 7 & 49 & 56 \\
& Total & 50 & 50 & 50 & 150\\
\end{tabular}
\caption{\label{confA2}The confusion matrix for  the  two shape ratios.}
\end{table}

  Alternative to consider are the shape variables based on the two areas 
 $$u=\text{sepal length} \times \text{sepal width},  v= \text{petal width} \times \text{petal width}=({x_1}{x_2} , {x_3}{x_4}).$$ 
 If we use $(u,v)$ as the variables,  it is found that  the   discriminant function is 
 $$  0.21 u -0.84 v. $$
 The confusion matrix  for the shape areas $(u,v)$  is given  in Table \ref{confArea} so there are 4 observations misclassified  with   1 observation of   versicolor go to virginica and 3 observations  of  virginica goes to  versicolor.
 
\begin{table}[htbp]
\begin{tabular}{ll|ccc|c}
&& \multicolumn{3}{c}{Actual}\\
&& setosa & versicolor & virginica & Total\\
\multirow{2}{*}{Predicted} & setosa & 50 & 0 & 0 & 50 \\
&  versicolor & 0 & 49 & 3 & 52 \\
&  virginica & 0 & 1 & 47 & 48 \\
& Total & 50 & 50 & 50 & 150\\
\end{tabular}
\caption{\label{confArea}The confusion matrix for the area shape variables  $(u,v).$  }
\end{table}

\noindent
Thus the area variables do a bit better than the Index variables but the genetic discriminant and the canonical discriminant are  marginally better.
These shape discriminants  have  the advantage of  using  the   two dimensional data summary so visualisation is simpler.

For completeness, let us go back to the two variables   $x_1$ sepal length and  $x_2$ sepal width we have  used in Section \ref{Topic2Classify}. It is found that the canonical  discriminant using these two variables  is 
$$ 2.14x_1 - 2.77 x_2 $$
so both $ x_1$  and  $x_2 $ contribute.  The confusion matrix is given in Table \ref{conf2sepal}
and it can be seen that the classification errors are higher (30 misclassified now) much higher  than the shape variables above  (10 misclassified) with these two variables which is consistent that we need all the four variables. 

\begin{table}[htbp]
\begin{tabular}{ll|ccc|c}
&& \multicolumn{3}{c}{Actual}\\
&& setosa & versicolor & virginica & Total\\
\multirow{2}{*}{Predicted} & setosa & 49 & 0 & 0 & 49 \\
&  versicolor & 1 &36& 15 & 52 \\
&  virginica & 0 & 14 & 35 & 56 \\
& Total & 50 & 50 & 50 & 150\\
\end{tabular}
\caption{\label{conf2sepal}The confusion matrix for the sepal length $x_1$ and sepal width $x_2$.}
\end{table}

\subsection {Testing of the Genetic  Hypothesis}\label{Test}
 From our discussion in the beginning of this section we say  that if  versicolor takes an intermediate value such that it differs twice as much  from
     setosa as from virginica then   the null  hypothesis from  (\ref{Interpoint})  is
$$ H_0: \bm {\mu_1} -3\bm {\mu_2}+2\bm {\mu_3}=0.$$

Substituting  the sample means from Table \ref{means}  into the genetic discriminant (\ref{u}),  we get  $\Bar{u}_1=-10.8$, $\Bar{u}_2=22.9$, $\Bar{u}_3=38.2$, so $\Bar{u}_1-3\Bar{u}_2+2\Bar{u}_3=-3.1$.
 Note that \citet{fisher1936}  denotes  the first population as virginica, second as versicolor and third as setosa, and    gives  more decimal points. He  gives 
 $\Bar{u}$=3.07052; var($\Bar{u}$)  =4.8365, SE($\Bar{u}$) =2.199. As $n$ is large, we could use  normal theory  so the 95
 ${\%}$ confidence interval is (-1.33,7.47) and  we accept the  null hypothesis. Fisher points out that this test is not exact and does the correction but the conclusion is the same.
  
\section{Visualisation}\label{Topic2Visual}
\begin{figure}[!htb]
\begin{center}
\includegraphics[scale=.07,angle=0]{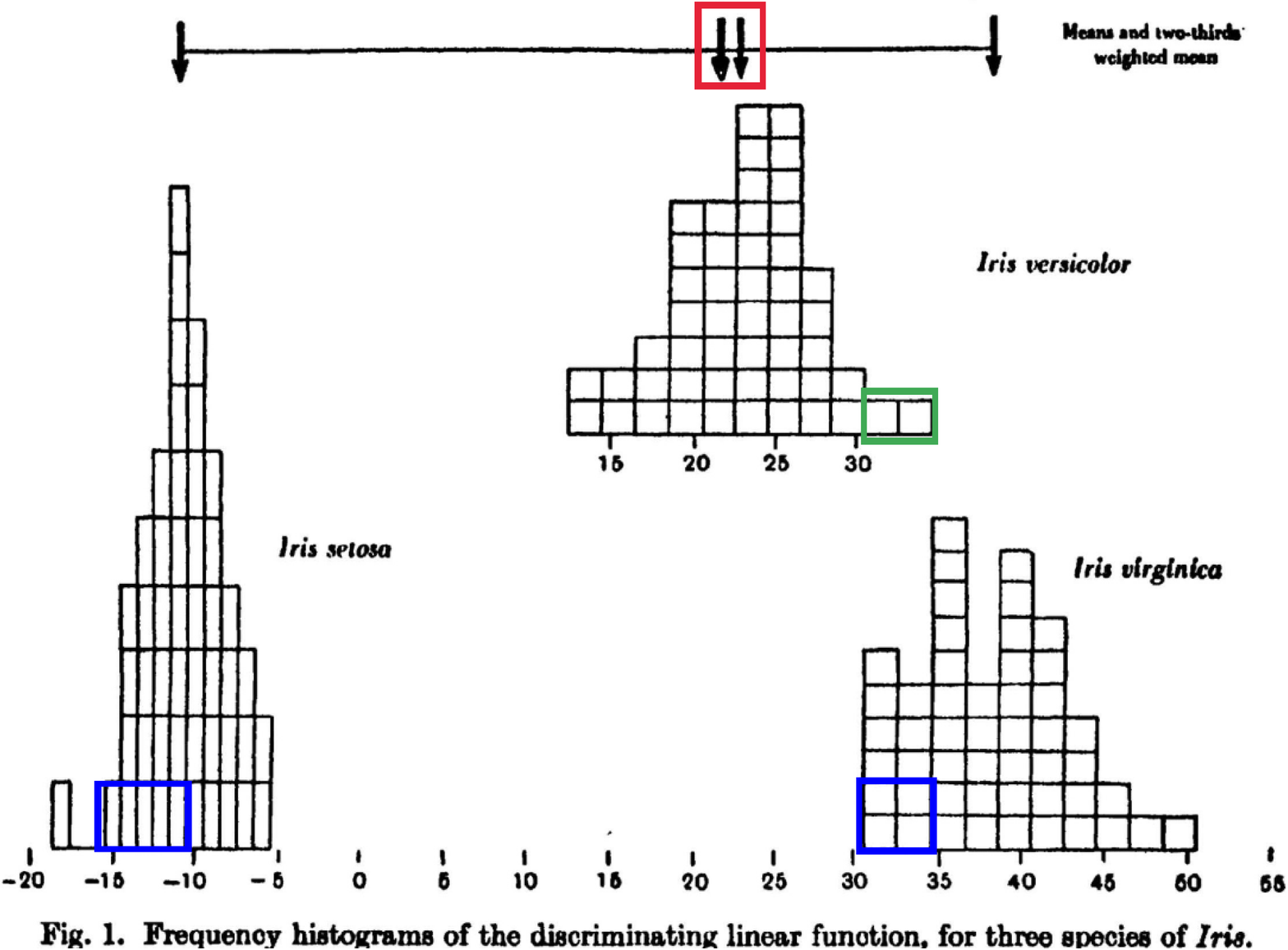}\\
\end{center}
 \caption{ \label{FFig1} Based on  \citet{fisher1936}, Figure 1: a  visual summary of the genetic discriminant for the three species via histograms. The top horizontal line shows each of the three sample means, and, additionally, the mean for versicolor  under the genetic assumption, thicker arrow. The text gives more details on the  boxes.}
\end{figure}

\citet{fisher1936}  found a way to visualise the iris data by drawing histograms of  his genetic discriminant  $\hat {u}$ given by  (\ref{u})  for the three populations (his Figure 1, reproduced here  as Figure \ref{FFig1}).

 We have added some annotations to  his  Figure 1 in our Figure \ref{FFig1} (coloured boxes)  as it depicts 4 important features as follows. (a) The  histogram of versicolor when compared to the histogram of virginica and seotsa, it can be seen that  there is overlap between the versicolor and virginica  histograms.  (b) There are just two observations of versicolor, in the green box on that histogram, that are not classified correctly, they are allocated to virginica, by the genetic discriminant. (c)  On the top horizontal line in the figure    the three  mean values and the mean value for versicolor under the genetic hypothesis are plotted (thicker arrow in the red box); versicolor takes an intermediate value  differing twice as much from  setosa as from virginica;  visual inspection indicates the plausibility of the null hypothesis.
(d) The sample standard deviations for the genetic discriminant are (2.4, 4.2, 4.3) for setosa, versicolor, and virginica respectively  and 
Fisher plots each cell of the histograms, corresponding to a single observation, of the "histogram" for setosa with half the width of the cells in the histograms for versicolor and virginiica.   Accordingly, to preserve area, the cells for setosa are each twice the height  as indicted by the blue boxes for setosa and virginca histograms  (versicolor and virginica have the same heights).

\begin{figure}[!htb]
\begin{center}
\includegraphics[width=0.7\linewidth]{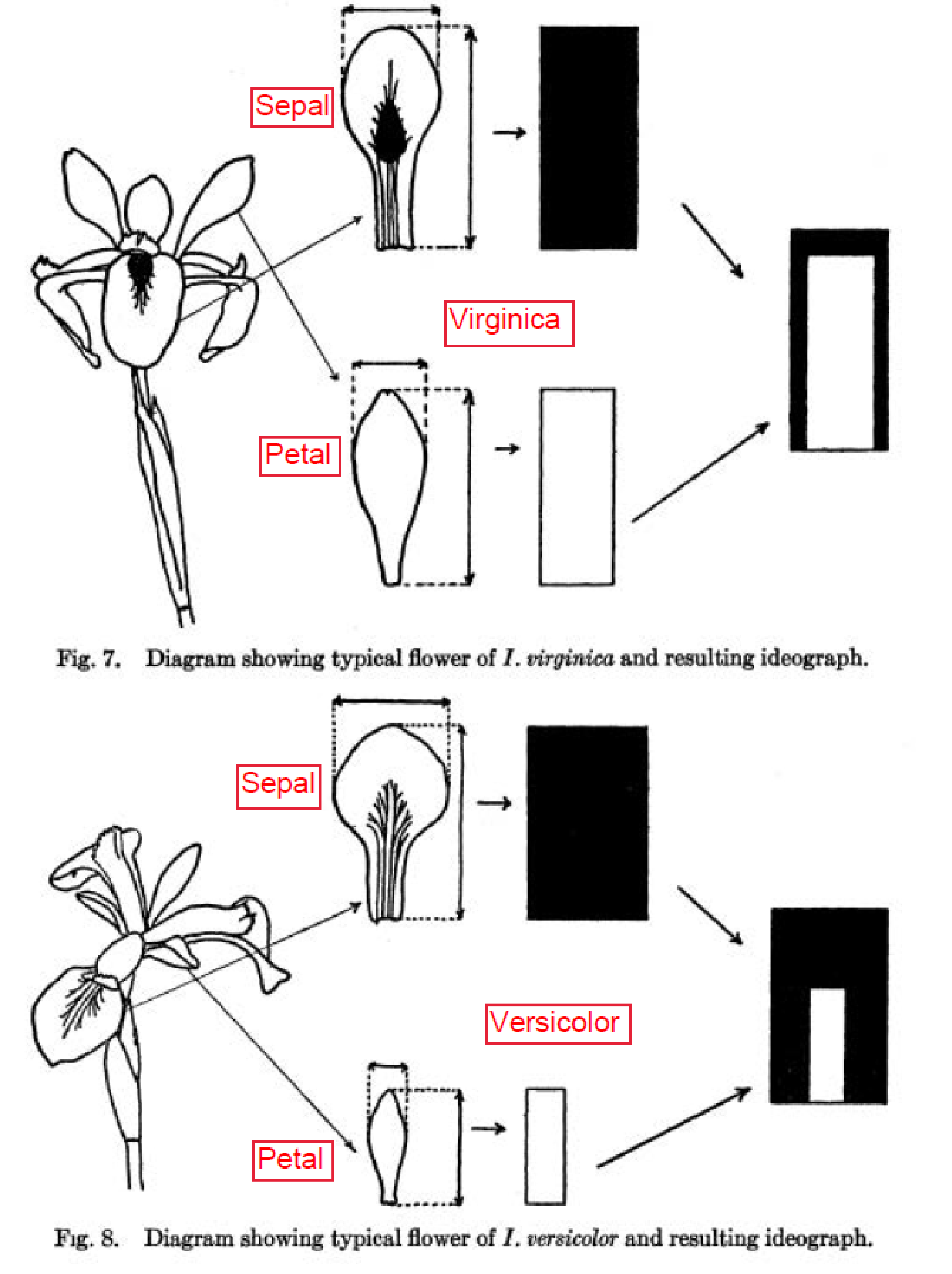}
\end{center}
 \caption{\label{ideog} E. Anderson's ideographs for typical  versicolor and virginica (reproduced from E.\citet{Anderson1928} page 285) . Red boxes added by the author.  }
\end{figure}

\begin{figure}[!htb]
\begin{center}
\includegraphics[width=1.0\linewidth]{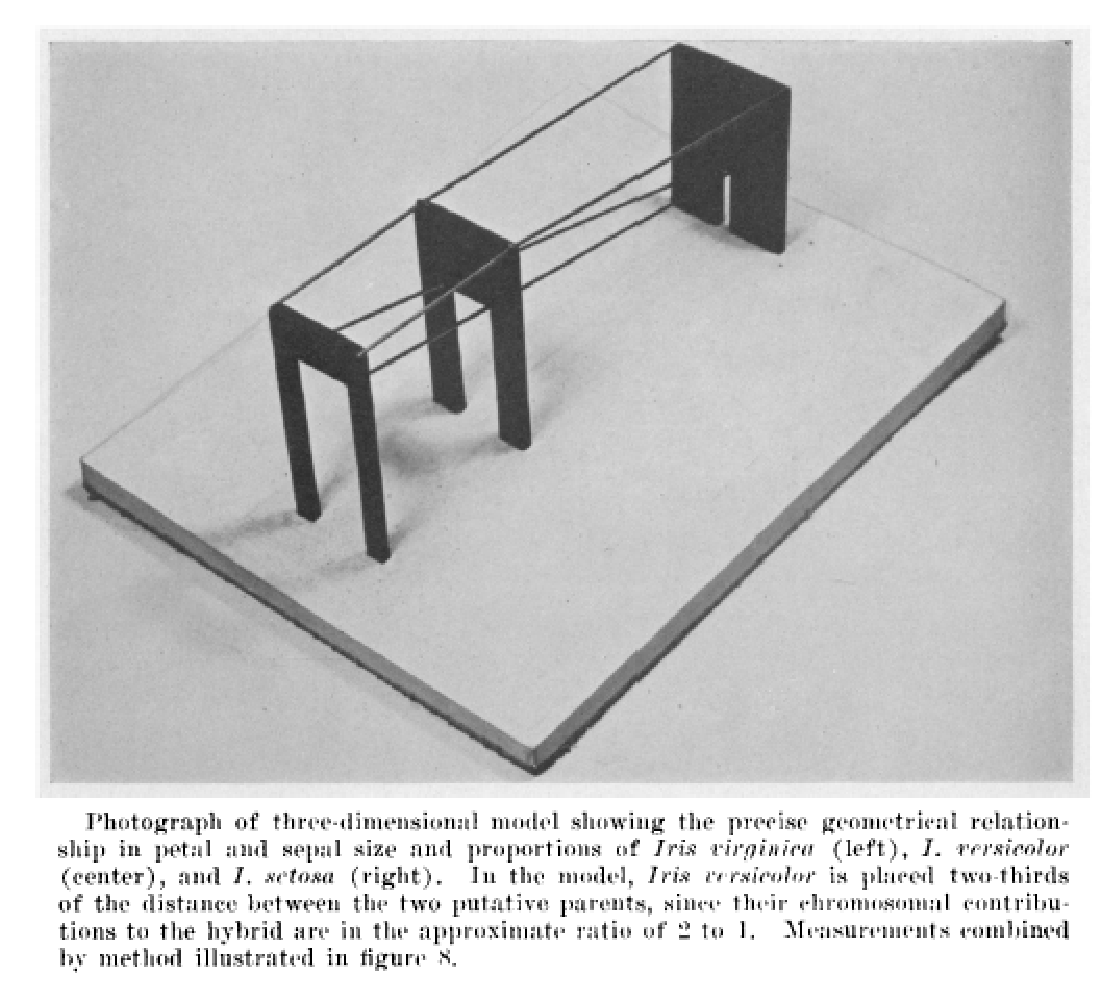}
\end{center}
 \caption{\label{AnderGenetic} E. Anderson's 3-dimensional model of ideographs for the iris data, depicting  the genetic hypothesis. Reproduced from \citet{Anderson1936}.}
\end{figure}

Interestingly,  E. \citet{Anderson1928}
 created what he called ``Ideographs''  which show all  four  variables at once by constructing a white rectangle with the length and the width for petal which is superimposed on a black rectangle with the same dimensions for  sepal.  Figure \ref{ideog} shows these for typical  virginica and versicolor observations. We can
see the  difference in comparing the two final rectangle that how the two species differs.  E. \citet{Anderson1928} was not only in touch with Fisher but also John Tukey and he took this   ideograph representation   further to allow for more features leading to 
``metroglyphs'' an extension of glyphs. For more details, see \citet{Klein2002}. Note that the  areas of the two  rectangles (bivariate data)  for visualisations of the full iris data have been used in a scatter plot  by \citet{Wain2001}.
\\
In fact, E \citet{Anderson1936}   created a 3-dimensional model of his ideographs for the iris data depicting his genetic hypothesis as reproduced in Figure \ref{AnderGenetic}. In the caption to this Figure, he refers to Figure  8 which is here   Figure \ref{ideog}.

\begin{figure}[!htb]
\begin{center}
\includegraphics[width=1.0\linewidth]{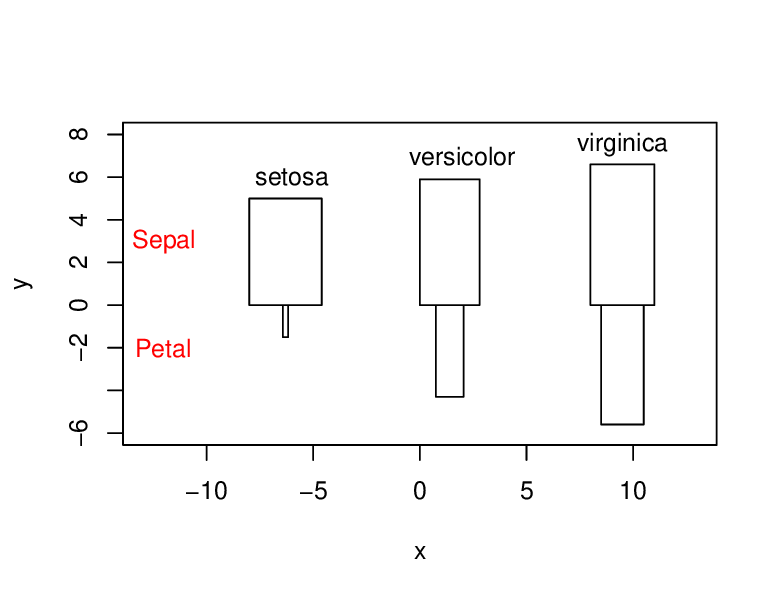}
\end{center}
 \caption{\label{Rect} Stacked-Rectangle  plot  of the mean vectors of setosa,  versicolor, and 
virginica. }
\end{figure}
\begin{figure}[!htb]
\begin{center}
\includegraphics[width=0.8\linewidth]{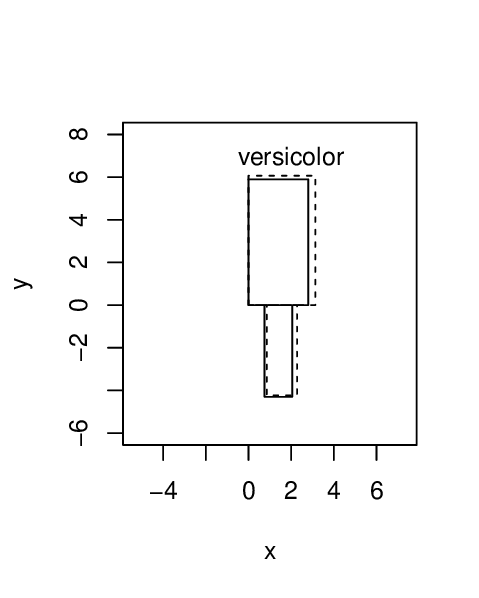}
\end{center}
 \caption{\label{HypVe} Stacked rectangles for versicolor and the genetic hypothesis. The observed versicolor means (solid lines) and  the observed   means from the   hypothesis (the dotted lines).}
\end{figure}

Following the concept of ideographs,  a simple presentation, an alternative to Figure \ref{ideog}, is to construct ``Stacked-Rectangle'' plots of the mean vectors for each iris species  with length along the  y-axis and width along the  x-axis. In Figure \ref{Rect} we give  such a plot with  sepal measurements prescribe the top rectangles, with rectangles for petal below, with a common edge. It can be seen that setosa is very different from versicolor and virginica, and  further  that virginica is ``bigger'' than versicolor.
Further, an alternative to the 3-dimensional model of the  genetic hypothesis  in  Figure \ref{AnderGenetic} of E. Anderson, 
we give our stacked rectangle version in Figure \ref{HypVe} using  the observed means of versicolor for the solid rectangle  and   using the means of setosa and virginica into (\ref{Interpoint})  computed for versicolor means under the hypothesis  for the dotted rectangle. We can see clearly the closeness between the observed means and  the ``hypothesis'' means for versicolor.

\begin{figure}[!htb]
\begin{center}
\includegraphics[width=1.0\linewidth]{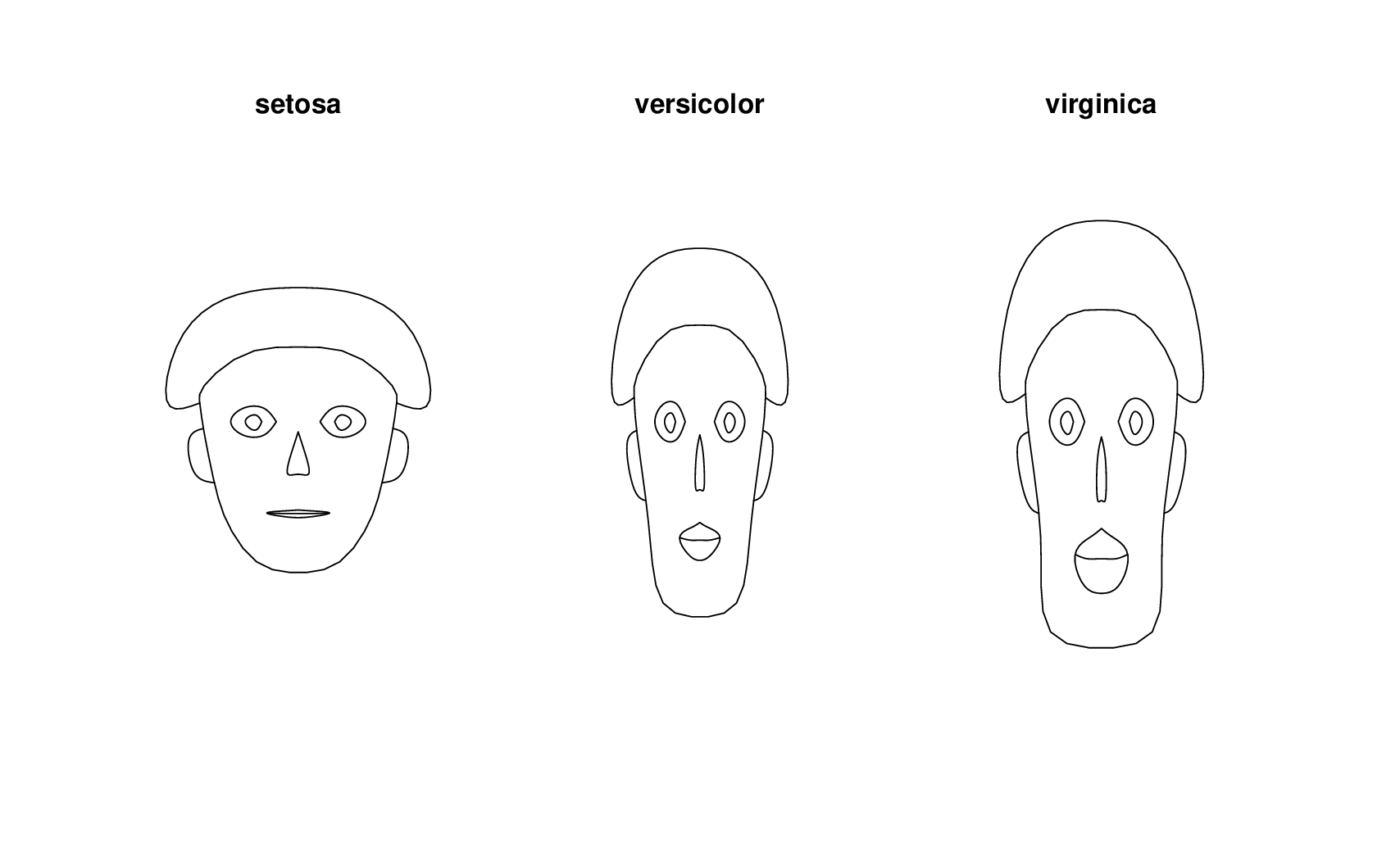}
\end{center}
 \caption{\label{Cfaces} Chernoff Faces corresponding to the mean vectors of setosa,  versicolor, and 
virginica. }
\end{figure}

Chernoff face representations have been popular (see for example, \citet{kvm1979}, Chapter 1).  Figure \ref{Cfaces} shows three faces corresponding
to the mean vectors of  setosa,  versicolor, and 
virginica;  plotted using the R library aplpack with all  four variables --  sepal length and sepal  width, petal length and petal width. Again, we note that the face associated
with setosa looks quite different from the other two
faces.

\begin{figure}[!htb]
\begin{center}
\includegraphics[scale=1.1,angle=0]{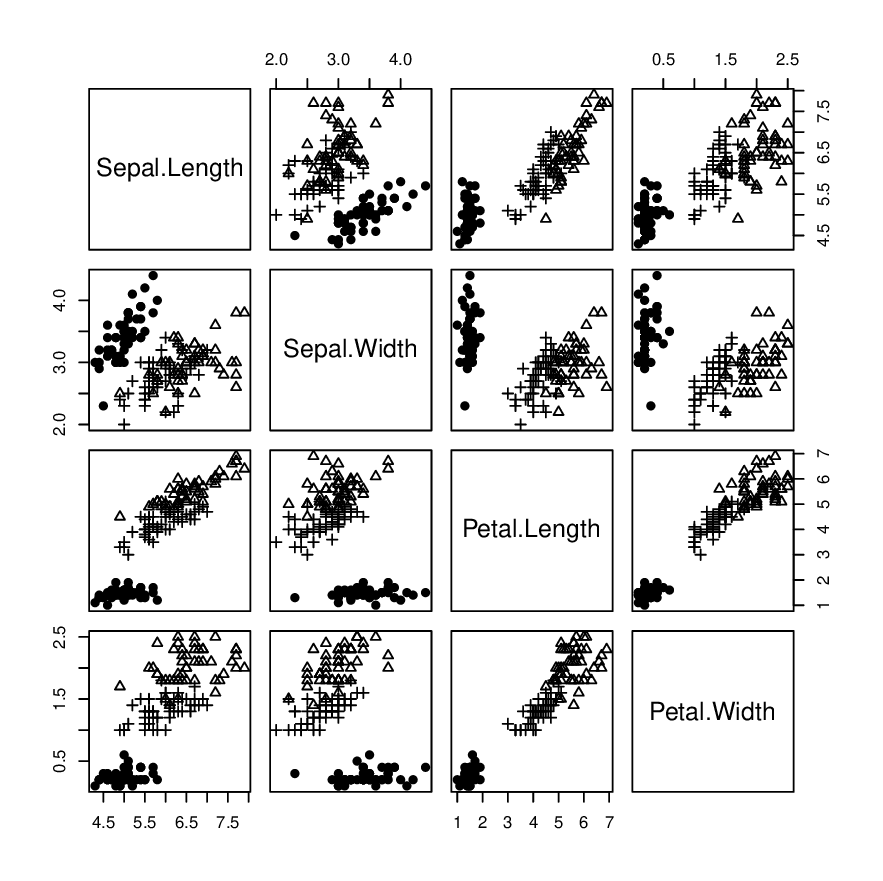}\\
 \end{center}
 \caption{ \label{ScatterP}Visualisation of the iris data: matrix scatter plots  (Symbols: $\bullet$  for setosa, $+$ for versicolor and $\triangle$ for  virginica).  }
\end{figure}

It has now become  common to draw  a set of 2D scatter plots  or even a set of  the perspective  3D  plots  in colour for higher dimensional data, both in print and utilizing visualization software.
For the iris data, Figure  \ref{ScatterP} gives a scatter plot matrix. It is clear that setosa has no overlap with the other two species,  but those  two species (versicolor and virginica) do  have  overlap and for some variables, the overlap is more pronounced than for others.

\begin{figure}[!htb]
\begin{center}
\includegraphics[scale=.9,angle=0]{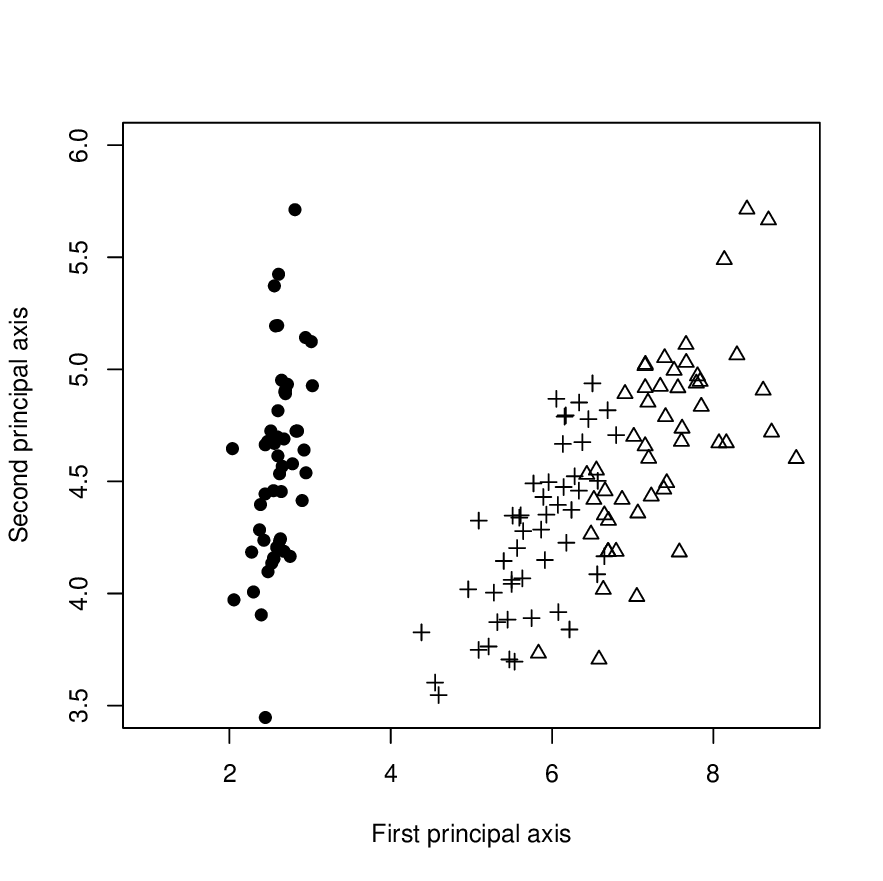}\\
 \end{center}
 \caption{ \label{4Dto2D}Visualisation of the iris data using Dhillon's method of principal components  (Symbols: $\bullet$   setosa,  $+$  versicolor and $\triangle$   virginica).  }
\end{figure}

Another approach is  using class preserving plots as developed by \citet{Dhillon2002},  which uses the first two principal components of the between sums of squares and cross products   matrix $B$.
Here, we have 

the first  principal component $( 0.327, -0.112, 0.863, 0.369 )$, and 

the second principal component $(-0.331, -0.888, 0.134, -0.288)$,

with  eigenvalues $(587.0, 5.1, 0, 0)$,

so all the information  is in  two dimensions.
 Figure \ref{4Dto2D} gives  the PCA plot.  In this one 2D  plot, one can visualise  the differences among the
three species more  vividly.

 We have already given a plot using  the first two canonical variables (using $W^{-1} B$) in   Figure \ref{Canonical}.
But  $W$, the within   sums of squares and cross products  matrix can be singular 
for high dimensional data so working with $B$ is preferred for visualisation.

In general, computing packages  have made the visualisation of multivariate data much easier,
including 2D and 3D scatter plots of raw data and derived components eg canonical variates,
and also visualization using interesting alternatives, e.g., Chernoff's faces and Andrews' function plots
(see \citet{kvm2023} for examples).
Still  there are visualisation tools described in the literature that have
yet to catch on, such as those described in the papers by 
\citet{tuk1981a}, \citet{tuk1981b} and  \citet{tuk1981c}.
However for printed publication, 2D representations (sometimes in colour) remain the most popular tool.

\section{A  Very Brief  History of the Appearance of Matrix Algebra in Multivariate Analysis }\label{MatrixH}

Matrix algebra is one of the most important mathematical  tools in statistics and in particular in multivariate analysis.
We give here a brief history, tracing the historical appearance  of matrix algebra  in the statistic literature.
Our treatment follows  \citet{david2006},  and the very recent  \citet{bingham2022} (seems to be written independently of the earlier  \citet{david2006} paper) .

Fisher relied on the power of $n$-dimensional geometry in his research work and avoided matrix algebra.
From the following comment  in   \citet{fisher1939}, his preference is clear.
{\it ``The paper incorporates the solution of the simultaneous distribution.
of the latent roots which arise in discriminant analysis, without
the formidable notation of matrix algebra. The method of resolution
of this rather difficult problem may therefore be of interest in
view of other possible applications."}

However, Fisher does use matrix algebra and matrix terminology.
Even in \citet{fisher1936},
the title of one of his tables reads {\it``Table IV. Matrix of multipliers reciprocal to the sums of squares and products within species ($cm^{-2}$)."}

\begin{figure}[!htb]
\begin{center}
\includegraphics[scale=1,angle=0]{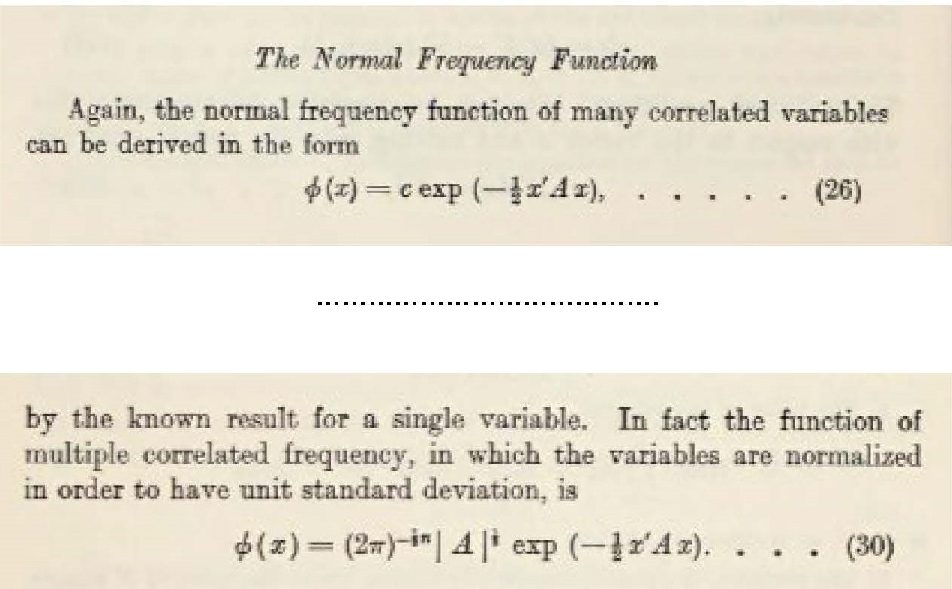}\\
 \end{center}
 \caption{\label{Aitken}The multivariate normal density given  in  \citet{turnbull1932}, pp. 174-175 ($A$ is now known as the precision/concentration  matrix and   $A^{-1}$  is the covariance matrix.)   }
\end{figure}

  According to  \citet{david2006}, the first appearance of matrix algebra in 
a statistics book seems to be in \citet{turnbull1932}.  They gave the multivariate normal density
in completely modern notation, eg Figure \ref{Aitken}.
According to  \citet{david2006}, credit for the introduction of matrix algebra into statistics 
must go to A. C. Aitken for the book cited and 
his earlier paper, \citet{aitken1931}. Surprisingly, Aitken did not use matrices in his book
``Statistical Mathematics'', \citet{aitken1939}, and later editions.

Matrix algebra begins to make fairly frequent appearances in  the 1940's, though not without some resistance.
\citet{cramer1946}  has a chapter introducing matrices which he then uses in subsequent chapters. 
Bartlett in his seminal discussion paper \citet{bartlett1947} writes, seemingly defensively,
{\it ``Perhaps I should add   that while  I have avoided any complicated analytical  discussion of theoretical problems,
I have not hesitated on occasion to refer to the mathematical theory,
with the aid of matrix and vector algebra or associated geometrical representation."}

During  my discussion of the autobiography of George Box published in 2013  \citet{box2013},
David Cox  mentioned  to me  that   Box in his BSc degree examination used  matrix algebra
to answer   a question on the  linear model and the examiner Florence David
gave him  poor marks for using matrix algebra, even though the answer was perfectly correct.
George Box obtained his  BSc degree  in mathematical statistics from University College London in 1947, and  Florence David  was a member of staff  of  the Statistics Department there from 1945-1962. 

 In the 1950's, matrix algebra became well established in multivariate analysis with the books of \citet{CRrao1952} and
 \citet{anderson1958}, and subsequently the extensive \citet{CRrao1965} (second edition \citet{CRrao1973}).
\citet{sen1987} points out in discussion to \citet{Scher1987}, itself a review of the second edition of Anderson's  book,
{\it ``It is undoubtedly true that a generation of (mathematical) statisticians
(specially, in the North American continent)
has been raised on the classical (1958) textbook of T. W. Anderson, An Introduction to Multivariate Statistical Analysis''}.

In fact, \citet{cramer1966} in their review of Multivariate Analysis give another assessment.
{\it ``The standard reference in multivariate analysis is undoubtedly Anderson's (1958) book,
An Introduction to Multivariate Analysis,
 but its high difficulty level and the paucity of examples make it an unsuitable reference for the research worker.
Rao's (1952) book remains an important reference for the research worker,
with its emphasis on applications of discriminant theory;
his later work (1965a) has some overlap in the multivariate analysis area but is at a higher mathematical level.''}  
Rao's (1965a)  cited here in this quote  is   \citet{CRrao1965}.

\begin{figure}[!htb]
\begin{center}
\includegraphics[scale=.5,angle=0]{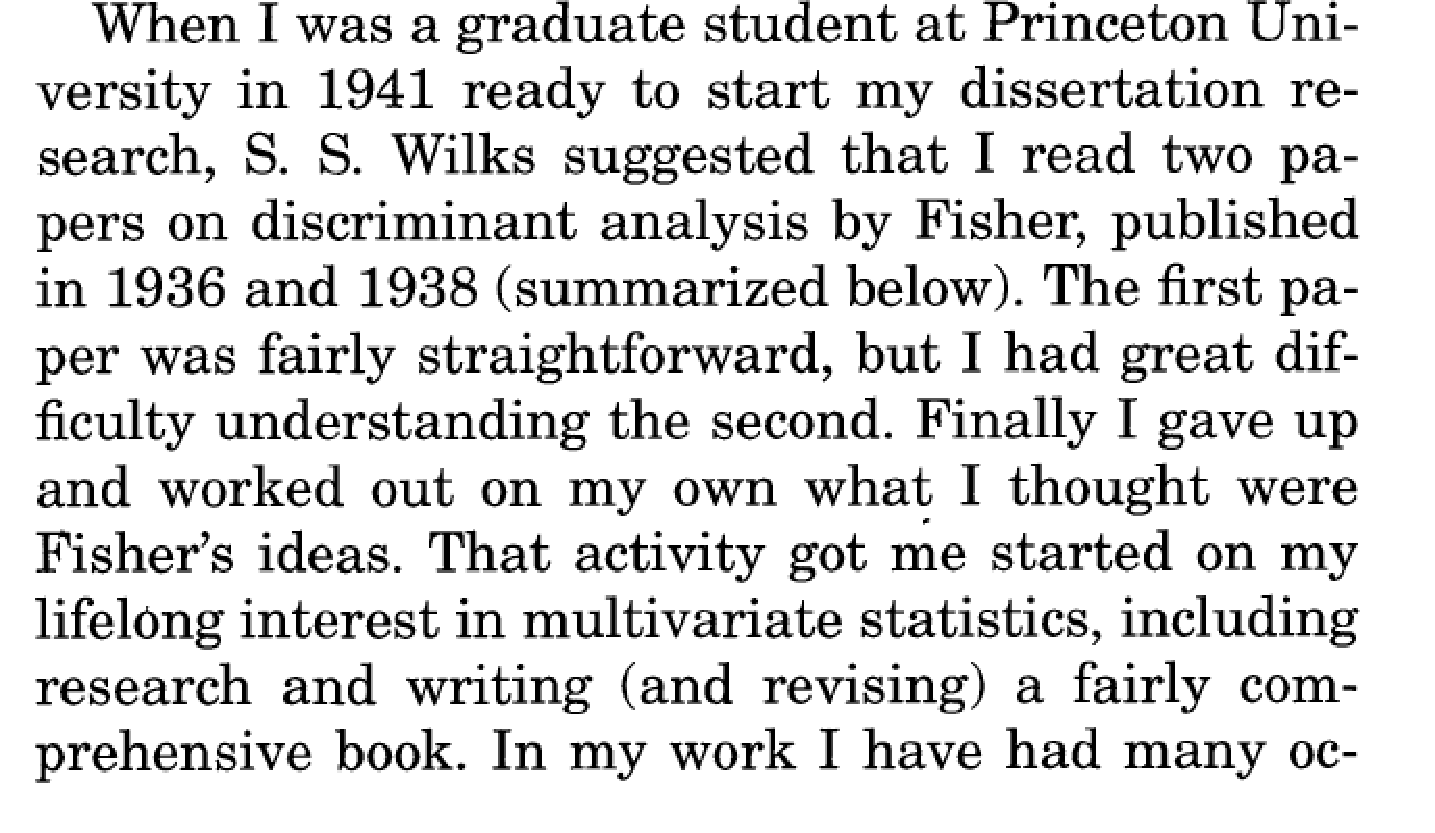}\\
\end{center}
 \caption{ \label{TWAnderson} The beginning  of T W Anderson  into Multivariate  Analysis  from Fisher's  discrimination papers; extracted from \citet {TWAnderson1996}. }
\end{figure}

Indeed,  the book by T.W. Anderson (\citet{anderson1958}) on Multivariate Analysis  (with various editions)  is a classic, and it is interesting to note that he started his work in the field as a Ph.D student reading the Fisher's papers on discriminant analysis (see his description in  Figure \ref{TWAnderson}).

On a personal note, in my two-year M.Sc. course in Statistics from Bombay University in 1955-1957
we used the celebrated Kendall's volumes, \citet{MGkendall1948} and
\citet{MGkendall1951}, as well as \citet{cramer1946} and \citet{CRrao1952}.
At that time the pioneering books devoted solely  on multivariate analysis,  \citet{MGkendall1957} and  \citet{anderson1958}, were not yet published.
Professor Kshirsagar was one of my teachers; subsequently he wrote his influential book \citet{Kshir1972}
in multivariate analysis.

\citet{bingham2022} point out that the crux for  the formulae for the multivariate normal density
lies with the papers by  Edgeworth in 1892-3,  and \citet{stigler86}, Chapter 9, pp 322-325, calls this Edgeworth's theorem.
However, the matrix notation for the multivariate density as used now  is due to Aitken as described above
though it is with the concentration matrix and  not the covariance matrix.
On the other hand, his joint book  (\citet{turnbull1932}) does derive the  second order moments.  
We refer for more details to the papers of \citet{david2006}  and \citet{bingham2022}.

\section{Statistical Learning and AI}\label {Learning}
Learning problems in multivariate analysis can broadly be classified into two groups,
 supervised learning and unsupervised learning.
In supervised learning, the goal is the same as  in   discriminant analysis
whereas in  unsupervised learning, the goal is the same as in cluster analysis.  
 More details on this topic  can be found in  \citet{EfronH2016},  \citet{Hastie2009}  and \citet{kvm2023}.  
In  modern techniques in these areas,   computation plays  a key role.
 Note that  discriminant analysis  is
also known as classification, pattern recognition,  machine learning  or statistical learning
though the emphasis might differ.

 In particular, to quote \citet{Sharma2013} {\it ``Pattern Recognition is one of the very important and actively searched trait or branch of artificial intelligence. It is the science which tries to make machines as intelligent as human to recognize patterns and classify them into desired categories in a simple and reliable way.''}
 Indeed, with recent overwhelming interest in AI, there have been general questions on how
 important is  the role of statistics. For example,\citet{Faes2022} have pointed out {\it  ``The research on and application of artificial intelligence (AI) has triggered a comprehensive scientific, economic, social and political discussion. Here we argue that statistics, as an interdisciplinary scientific field, plays a substantial role both for the theoretical and practical understanding of AI and for its future development. Statistics might even be considered a core element of AI.'' }
Also, it is to be noted that the  terms used in the statistical vs. machine learning/AI world are not in general the same and \citet{Faes2022} have provided a table giving  one to one correspondence between the two areas. \citet{Ghahr2015}  has also given an excellent review on statistical learning  
and artificial intelligence.

We have already described Fisher's linear discriminant
function. It can be viewed as an intuitive
approach to discriminant analysis that looks for a ``sensible''
rule to discriminate between populations.  It relies on the means and covariances, and one of  its extensions  has been to take the  populations as normal.

Alternative nonparametric methods are also available;
some are given in Section \ref{Topic2Classify}.  The simplest of these methods, $k$-nearest neighbour,
allocates a new observation $\bm x_0$
according to a majority vote among the labels of the $k$ nearest
neighbours in the data to $\bm x_0$.

Other nonparametric methods use recursive partitioning. As the name suggests, this method
successively divides the data into subdivisions.
Each partition corresponds to a rule based on an inequality for a one-dimensional
function of the attribute data ${\bm  x}$, 
for example, ${\bm  a}^T {\bm  x} \ge 0$.  The choice of partition depends on the purity (which 
uses the associated labels) of the data in each resulting subdivision.
We can describe this process by a sequence of rules,
usually based on components of $\bm x$, and these can be represented as 
a classification (or decision) tree.
Some recent approaches, for example ensemble methods, boosting, and random forests, are closely connected to 
classification trees.  Many of such methods are more algorithmic than model-based.    

Logistic regression also makes few 
assumptions about the distribution of the data. 
Some recent approaches include neural 
network / deep learning families of algorithms, utilizing
 multi-layer perceptrons, of which logistic regression represents a simple case,
 radial basis functions and support vector machines.  These have all been 
widely used for discriminant analysis. 

 However, the area of machine / statistical learning  is very broad,
 and extends beyond multivariate analysis to many fields of statistics,
 such as spatial analysis (see \citet{Kentmardia2022}).
 The original motivation for neural networks to mimic the possible learning
behaviour of the human brain is now considered part of neuroscience in AI, connected
by statisticians to  discriminant analysis.   It seems Brian Ripley was the first to make that connection
 in a series of papers (see, for  example, \citet {Ripley1994}).  In particular  the iris data, as an early example of  illustrating  neural networks, has been used in  \citet {Ripley1994}.

It is important to note that the aim of discriminant analysis is to find a rule  to separate distinct populations, and then to use  that rule (or a different classification scheme) to  allocate a new observation  to the given populations. \citet {fisher1936} has both these elements.
In the initial sections he developed  how to form a discriminant rule and subsequently,  in his Figure \ref{FFig1}, he carried out his classification using the genetic discriminant. Of course, that is just one method to  carry out classification.

\cite {Friedrich2022} have stressed that 
{\it ``The objective of statistics related to AI must be to facilitate or enable the interpretation
of data. As Pearl puts it: ‘Data alone are hardly a science, regardless how big
they get and how skillfully they are manipulated’ (\citet {Pearl2018}). What is important is
the knowledge gained that will enable future interventions.''}  We conclude with the following important message in \citet{Faes2022} {\it ``Fundamental work
on ML goes back to the 1960s and was developed on
the basis of mathematical and statistical principles to which
traditional statistics also refers (\citet {Foote2021}). It is worth remembering
its roots.''}
(Here ML  stands for Machine Learning.)

Remarkably, one of the leading pioneers in AI, Stephen Wolfram, has summarised his vision succinctly in  \citet {Wolfram2023}:
{\it ``For decades there’s been a dichotomy in thinking
about AI between “statistical approaches” of the kind
ChatGPT uses, and “symbolic approaches” that are in
effect the starting point for Wolfram$|$Alpha. But now
—thanks to the success of ChatGPT—as well as all
the work we’ve done in making Wolfram$|$Alpha
understand natural language—there’s finally the
opportunity to combine these to make something
much stronger than either could ever achieve on their
own.''}

 Undoubtedly, the Fisher Rule is the first  statistical rule  for discrimination. It plays  a major role in  supervised learning  and is the foundation  more generally of  AI. 

\section{Acknowledgements}
 The author is grateful to  Colin Goodall, John Kent, Wojtek Krzanowski,  Charles Taylor and  Xiangyu Wu for their helpful  comments and to the Fisher Memorial Trust for  inviting me to give the talk. Thanks are also due to the Leverhulme  Trust for the Emeritus Fellowship.

\bibliographystyle{rss}

\end{document}